\documentclass[12pt]{article}

\usepackage{latexsym}
\usepackage{amsmath}
\usepackage{amsfonts}
\usepackage{amssymb}
\usepackage{cite}
\usepackage[dvips]{graphicx}

\newcommand{\cK}{{\cal K}}
\newcommand{\cL}{{\cal L}}
\newcommand{\cM}{{\cal M}}
\newcommand{\cN}{{\cal N}}

\newcommand{\cS}{{\cal S}}

\newcommand{\beq}{\begin{equation}}
\newcommand{\eeq}{\end{equation}}
\newcommand{\bi}{\begin{itemize}}
\newcommand{\ei}{\end{itemize}}
\newcommand{\bea}{\begin{eqnarray}}
\newcommand{\eea}{\end{eqnarray}}
\newcommand{\ba}{\begin{array}}
\newcommand{\ea}{\end{array}}
\newcommand{\bt}{\begin{tabular}}
\newcommand{\et}{\end{tabular}}
\newcommand{\bc}{\begin{center}}
\newcommand{\ec}{\end{center}}

\newcommand{\Ga}{\alpha}

\newcommand{\GD}{\Delta}

\newcommand{\GTh}{\Theta}

\newcommand{\ft}[2]{{\textstyle {\frac{#1}{#2}} }}
\newcommand{\vl}{{\vphantom{[}}}

\newcommand{\ga}[3]{\Gamma^{#1}{}_{#3}{}^{#2}}
\newcommand{\gaquer}[3]{\overline{\Gamma}\vphantom{\Gamma}^{#1}{}_{#3}{}^{#2}}
\newcommand{\li}[1]{\hat{#1}}
\newcommand{\f}[1]{\boldsymbol{#1}}

\makeatletter
\@addtoreset{equation}{section}
\makeatother
\renewcommand{\theequation}{\thesection.\arabic{equation}}

\begin{document}

\begin{titlepage}
\begin{center}

\hfill hep-th/0503088\\
\hfill DESY 05-041\\
\hfill ZMP-HH/05-05

\vskip 1.5cm 
\mathversion{bold}
{\Large \bf Effective Actions for Massive\\[.5ex] 
Kaluza-Klein States on $AdS_3\times S^3\times S^3$
\\[0.2cm]}
\mathversion{normal}

\vskip 1.5cm

{\bf Olaf Hohm and Henning Samtleben} \\

\vskip 20pt

{\em II. Institut f\"ur Theoretische Physik\\ 
Universit\"at Hamburg\\
Luruper Chaussee 149\\
D-22761 Hamburg, Germany}\\

\vskip 15pt

{email: {\tt olaf.hohm@desy.de, henning.samtleben@desy.de}} \\

\vskip 0.8cm

\end{center}

\vskip 15mm

\begin{center} {\bf ABSTRACT}\\[3ex]

\begin{minipage}{13cm}
\small
We construct the effective supergravity actions
for the lowest massive Kaluza-Klein states on the 
supersymmetric background $AdS_{3}\times S^{3}\times S^{3}$.
In particular, we describe the coupling of the 
supergravity multiplet to the lowest massive
spin-$3/2$ multiplet $(\ft12,\ft12;\ft12,\ft12)_{{\rm S}}$
which contains 256 physical 
degrees of freedom and includes the moduli 
of the theory.
The effective theory is realized as the broken phase
of a particular gauging of the maximal
three-dimensional supergravity with gauge group 
$SO(4)\times SO(4)$. 
Its ground state breaks half of the 
supersymmetries leading to 8~massive gravitinos
acquiring mass in a super Higgs effect.
The holographic boundary theory realizes the 
large ${\cal N}=(4,4)$ superconformal symmetry.

\end{minipage}
\end{center}
\noindent

\vfill

March 2005

\end{titlepage}

\tableofcontents

\section{Introduction}
Among the celebrated AdS/CFT 
dualities~\cite{Maldacena:1997re}, 
the correspondence between 
two-dimensional conformal field theories and string theories on 
$AdS_3$ is of particular importance
not at least due to the infinite dimensional structure
of the conformal group.
Originally, the interest was mainly focused on string theory 
on $AdS_3\times S^3\times M_4$, with $M_4=K3$ or 
$T^4$ arising as the near-horizon geometry of the D1-D5 
system~\cite{Maldacena:1998bw}.
The dual conformal field theory is described by a 
non-linear $\sigma$-model whose target space is a 
deformation of the symmetric product orbifold
{\sf Sym}$^{N}(M_4)$. 
More recently, the focus has also turned to the duality
involving string theory on 
$AdS_3\times S^3\times S^3\times 
S^{1}$~\cite{Elitzur:1998mm,deBoer:1999rh,Gukov:2004ym,Gukov:2004fh,Lu:2002kw,
Sommovigo:2003kd}.
This geometry is half maximally supersymmetric and 
arises in the near-horizon limit
of the so-called double D1-D5 
system~\cite{Cowdall:1998bu,deBoer:1999rh}.
Its isometries form two copies of the 
supergroup $D^1(2,1;\alpha)$.
The ratio of brane charges $\alpha$ here coincides with the ratio of
the two sphere radii. 
Correspondingly, the dual
conformal field theory should realize the {\em large} 
${\cal N}=4$ superconformal 
algebra~${\mathcal A}_{\gamma}$.\footnote{In contrast 
to the small ${\cal N}=4$ superconformal algebra realized
by the boundary theory of the single D1-D5 system,
the large ${\cal N}=4$ algebra~${\mathcal A}_{\gamma}$ contains
two affine $\widehat{\mathfrak{su}(2)}$ factors, related to the 
isometries of the two $S^{3}$ spheres.}
It has been conjectured to be related to
a deformation of the
$\sigma$-model whose target space
is the symmetric product 
{\sf Sym}$^{N}(U(2))$~\cite{Elitzur:1998mm,deBoer:1999rh}.
Despite the larger symmetry, this holographic
duality is still far less understood than the 
case of the single D1-D5 system.

In the supergravity limit, 
the Kaluza-Klein (KK) modes on the
$AdS_3\times S^3\times M_4$ geometry 
are dual to chiral primary
operators in the conformal field theory. Although 
CFT calculations
have been mainly performed at the orbifold
point, where the supergravity approximation
breaks down, nontrivial tests of the dualities are possible for
quantities protected by non-renormalization theorems; in particular,
BPS spectra and elliptic genera were matched successfully
\cite{deBoer:1998ip}. From
the supergravity side this essentially requires linearization of the ten-dimensional
field equations around the $AdS_{3}$ background.
Knowledge of the full nonlinear three-dimensional
effective supergravity theory on the other hand
contains considerable information
about the conformal field theory
even beyond the level of the spectrum.
For instance, the full nonlinear couplings
of the supergravity fields encode the 
higher order correlation functions on the CFT side.
Furthermore, in order to analyze renormalization group flows 
in the field theory via the dual supergravity description, 
one needs the full scalar potential of the supergravity theory
which in particular encodes information about the infrared
fixpoints of these flows~\cite{Berg:2001ty}.  

For the case of $AdS_{3}\times S^{3}$, the full effective supergravity theory
has been constructed in~\cite{Nicolai:2003ux},
drawing on the special properties of gauged supergravities in three 
dimensions~\cite{Nicolai:2000sc,Nicolai:2001ac,Nicolai:2003bp,Lu:2003yt,deWit:2003ja}.
The interactions of the infinite Kaluza-Klein towers
of massive spin-$1$ multiplets
with the massless ${\cal N}=8$ supergravity multiplet have
been described in terms of a gauged 
supergravity over a single irreducible coset space.
It comes with a local $SO(4)$ gauge symmetry
related to the isometries of the $S^{3}$ sphere ---
and in the holographic context to the $R$-symmetry
of the conformal field theory.
The spin-$1$ fields acquire mass in a three-dimensional variant of the
Brout-Englert-Higgs mechanism
involving an infinite number of fields.
An open problem has remained the inclusion of 
the infinite tower of massive spin-2 multiplets
in this analysis.

On the $AdS_{3}\times S^{3}\times S^{3}$ background,
the Kaluza-Klein spectrum organizes into infinite towers
of massive spin-$3/2$ and massive spin-2 multiplets.
Massive spin-$3/2$ fields can be realized through spontaneously
broken local supersymmetry 
and we can therefore be optimistic that at least 
some part of this Kaluza-Klein tower can be incorporated into a 
supergravity description.
In this paper we will focus on the lowest multiplets appearing 
in the KK spectrum. 
In particular, we will construct the effective theory that 
describes the interactions between the
massless supergravity multiplet and
the lowest massive spin-3/2 multiplet 
$(\ft12,\ft12;\ft12,\ft12)_{\rm S}$.\footnote{For the
notation of supermultiplets we follow \cite{deBoer:1999rh}
and denote by $(\ell_{L}^{+},\ell_{L}^{-};\ell_{R}^{+},\ell_{R}^{-})_{\rm S}$
the short supermultiplet generated from the highest weight state
$(\ell_{L}^{+},\ell_{L}^{-};\ell_{R}^{+},\ell_{R}^{-})$ where
$\ell_{L,R}^{\pm}$ denote the spins under the various $SU(2)$ factors of $SO(4)\times SO(4)$,
see section 2.1 for details.}

This massive multiplet contains 256 physical degrees of freedom,
among them eight massive spin-$3/2$ fields. It 
is of particular interest for several reasons. 
First of all, this multiplet is the only one appearing in
the Kaluza-Klein spectrum that carries a massless scalar
field which is invariant under the full $SO(4)\times SO(4)$.
These fields correspond to the moduli of the theory, i.e.~they
are dual to marginal operators of the conformal field theory.
Second, the massive spin-$3/2$ multiplet is
the lowest massive multiplet 
of the particular 
type $(\ell_{L},\ell_{L};\ell_{R},\ell_{R})_{\rm S}$,
which implies that its conformal dimensions are protected
throughout the moduli space.
In contrast, generic multiplets appearing in the Kaluza-Klein tower
are expected to receive quantum corrections~\cite{deBoer:1999rh, Gukov:2004ym}. 
This is a peculiarity of the theories with large ${\cal N}=4$ symmetry,
as the BPS condition of the underlying superconformal 
algebra ${\cal A}_{\gamma}$ is nonlinear.
Finally, together with the supergravity multiplet,
the multiplet $(\ft12,\ft12;\ft12,\ft12)_{\rm S}$ sits in the same
short representation of the superconformal algebra ${\cal A}_{\gamma}$.
The combined theory may thus be relevant to
realize the full ${\cal A}_{\gamma}$ symmetry.

In this paper we will construct the effective three-dimensional
supergravity theory describing the interactions of this massive multiplet
as a particular gauging of the maximal ${\cal N}=16$ theory
in its broken phase.
In particular, we give the complete scalar potential
of the theory as a function
on the coset manifold $E_{8(8)}/SO(16)$ which yields the 
correct KK scalar, vector and gravitino 
masses through a supersymmetric Brout-Englert-Higgs mechanism.
As a by-product this construction yields a new example of a maximal gauged 
supergravity theory within the classification 
initiated in \cite{Nicolai:2000sc,Fischbacher:2003yw}.
Interestingly, and in contrast to previously known examples,
it does not possess a symmetric phase but its ground state
spontaneously breaks half of the supersymmetries
down to an ${\cal N}=(4,4)$ supersymmetric $AdS$ geometry;
correspondingly eight of the gravitinos acquire mass.

This paper is organized as follows. 
In section~2 we review the Kaluza-Klein
spectrum on $AdS_3\times S^3\times S^3$,
as it has been computed in~\cite{deBoer:1999rh}. 
We describe in detail the lowest multiplets appearing
in the infinite Kaluza-Klein tower. 
The effective three-dimensional theories describing the 
couplings of the lowest massive spin-1/2 and spin-1 multiplets
to the massless ${\cal N}=8$ supergravity multiplet
can be constructed 
along the lines of~\cite{Nicolai:2001ac,Nicolai:2003ux}
as particular gauged ${\cal N}=8$ supergravities.
Particular attention is paid to the 
massive spin-3/2 multiplet $(\ft12,\ft12;\ft12,\ft12)_{\rm S}$.
We argue that its field content 
suggests the description as a particular 
gauged maximal ${\cal N}=16$ supergravity theory. 

Section~3 then presents the detailed construction of this theory.
We start with a brief summary of the general three-dimensional maximal gauged 
supergravity theory in section~3.1. The generic theory may 
be described as a deformation
of the maximal ungauged theory with global $E_{8(8)}$ symmetry.
The deformation is completely specified by the choice 
of the so-called embedding tensor describing the
minimal couplings between vector and scalar fields,
which in turn is subject to a set of algebraic consistency constraints.
We identify the relevant embedding tensor 
in section~3.3 and verify that
it is a solution to all the algebraic constraints.
In section~3.4 we prove the existence of a ground state
in this theory
which spontaneously breaks half of the supersymmetries
and exhibits the correct background isometry group 
$D^1(2,1;\alpha) \times D^1(2,1;\alpha)$. 
We compute the mass spectrum by linearizing the field equations
around this ground state and show that it
indeed coincides with the 
prediction from the Kaluza-Klein analysis.
As a first application, we compute
in section~3.5 the scalar 
potential for the scalar fields which
are singlets under the gauge group.
In particular, this confirms the uniqueness of the ground state.
In section~4 we outline future directions of work.

\section{Kaluza-Klein Supergravity on $AdS_3\times S^3\times S^3$}

\subsection{Supergravity spectrum}
We start by reviewing the Kaluza-Klein spectrum of ten-dimensional
supergravity compactified on $AdS_3\times S^3\times S^3\times S^{1}$
following~\cite{deBoer:1999rh}. As we will not discuss the higher 
Kaluza-Klein modes on the $S^1$, this corresponds to effectively starting 
from nine-dimensional supergravity. 
The spectrum can then be derived by purely group-theoretical methods.

The background isometry supergroup under which the spectrum organizes 
is the direct product of two ${\cal N}=4$ supergroups
\bea
D^1(2,1;\alpha)_{L} \times D^1(2,1;\alpha)_{R}
\;,
\label{supergroup}
\eea
in which each factor combines a bosonic 
$SO(3)\times SO(3)\times SL(2,\mathbb{R})$
with eight real supercharges~(see~\cite{Sevrin:1988ew}
for the exact definitions).  
More precisely, the noncompact factors
$SL(2,\mathbb{R})_{L}\times SL(2,\mathbb{R})_{R} = SO(2,2)$
join into the isometry group of $AdS_3$ while the compact factors
\bea
G_{\rm c}&=&
SO(3)_{L}^{+}\times SO(3)_{L}^{-} \times SO(3)_{R}^{+}\times SO(3)_{R}^{-}
\;,
\label{so34}
\eea
build the isometry groups $SO(4)^{\pm}\equiv SO(3)_{L}^{\pm} \times SO(3)_{R}^{\pm}$
of the two spheres $S^{3+}\times S^{3-}$.
Accordingly, this group will show up as the gauge group of the 
effectively three-dimensional supergravity action.
The parameter $\alpha$ of~(\ref{supergroup})
describes the ratio of the radii of the two spheres $S^3$.

The spectrum of the three-dimensional
supergravity theory combines into supermultiplets of the 
group~(\ref{supergroup}).
A short $D^1(2,1;\alpha)$ supermultiplet is defined by its highest
weight state $(\ell^+,\ell^-)^{h_{0}}$, where $\ell^{\pm}$ label spins
of $SO(3)^{\pm}$ and $h_{0}=\frac1{1+\alpha}\,\ell^{+}+ \frac\alpha{1+\alpha}\,\ell^{-}$
is the charge under the Cartan subgroup ~$SO(1,1)\subset SL(2,\mathbb{R})$.
The corresponding supermultiplet which we will denote by
$(\ell^+,\ell^-)_{\rm S}$ is
generated from the highest weight state by the action of
three out of the four supercharges $G^a_{-1/2}$ ($a=1,...,4$) and 
carries $8(\ell_{+}+\ell_{-}+4\ell_{+}\ell_{-})$ degrees of freedom.
Its $SO(3)^{\pm}$ representation content is summarized in table~\ref{short}.

\begin{table}[b]
\centering
  \begin{tabular}{c | c c c}
  $h$ & \\
  \hline
   $h_{0}$ & & $(\ell^+,\ell^-)$ & \\[.5ex]
   $h_{0}+\frac{1}{2}$ & $(\ell^+-\frac{1}{2},\ell^- -\frac{1}{2})$ &
   $(\ell^+ +\ft12,\ell^- -\ft12)$ & $(\ell^+ -\ft12,\ell^- +\ft12)$ \\[.5ex]
   $h_{0}+1$ & $(\ell^+,\ell^- -1)$ &
   $(\ell^+ -1,\ell^-)$ & $(\ell^+,\ell^-)$ \\[.5ex]
   $h_{0}+\ft32$ & & $(\ell^+ -\ft12,\ell^- -\ft12)$ &
  \end{tabular}
  \caption{\small The generic 
  short supermultiplet $(\ell^+,\ell^-)_{\rm S}$ of $D^1(2,1;\alpha)$,
with  $h_{0}=\frac1{1+\alpha}\,\ell^{+}+ \frac\alpha{1+\alpha}\,\ell^{-}$.} 
  \label{short}
\end{table}  

The generic long multiplet $(\ell^+,\ell^-)_{\rm long}$ 
instead is built from the action
of all four supercharges $G^a_{-1/2}$ 
on the highest weight state and %correspondingly 
carries 
$16\,(2\ell_{+}\!+\!1)(2\ell_{-}\!+\!1)$ degrees of freedom.
Its highest weight state satisfies the unitarity bound 
$h\ge\frac1{1+\alpha}\,\ell^{+}+ \frac\alpha{1+\alpha}\,\ell^{-}$.
In case this bound is saturated, the long multiplet decomposes
into two short multiplets (table \ref{short}) according to
\bea
(\ell^+,\ell^-)_{\rm long} &=& 
(\ell^+,\ell^-)_{\rm S} \oplus (\ell^+\!+\ft12,\ell^-\!+\ft12)_{\rm S}
\;.
\label{long}
\eea
The lowest short supermultiplets $(0,\ft12)_{\rm S}$,
$(0,1)_{\rm S}$, and $(\ft12,\ft12)_{\rm S}$ of 
$D^1(2,1;\alpha)$ are further degenerate and collected in table~\ref{lower},
and similar for $\ell^+\leftrightarrow\ell^-$, $\Ga \leftrightarrow1/\Ga$. 

\begin{table}[bt]
\centering
  \begin{tabular}{c|c}
   $h$ & $(\ell^+,\ell^-)$ \\ \hline
   $\frac{\Ga}{2(1+\Ga)}$ & $(0,\ft12)$ \\
   $\frac{2\Ga+1}{2(1+\Ga)}$ & $(\ft12,0)$ \\ \vl \\ \vl 
  \end{tabular}
 \qquad\quad
  \begin{tabular}{c|c}
   $h$ & $(\ell^+,\ell^-)$ \\ \hline
   $\frac{\Ga}{1+\Ga}$ & $(0,1)$ \\
   $\frac{3\Ga+1}{2(1+\Ga)}$ & $(\ft12,\ft12)$ \\ 
   $\frac{2\Ga+1}{1+\Ga}$ & $(0,0)$  \\ \vl 
  \end{tabular}
 \qquad\quad
  \begin{tabular}{c|c}
   $h$ & $(\ell^+,\ell^-)$ \\ \hline
   $\ft12$ & $(\ft12,\ft12)$ \\
   $1$ & $(0,0)+(0,1)+(1,0)$ \\
   $\ft32$ & $(\ft12,\ft12)$ \\
   $2$ & $(0,0)$   
  \end{tabular}
    \caption{\small The lowest short supermultiplets $(0,\ft12)_{\rm S}$,
$(0,1)_{\rm S}$, and $(\ft12,\ft12)_{\rm S}$ of  $D^1(2,1;\alpha)$.} 
\label{lower}
\end{table}

Short representations of the full supergroup~(\ref{supergroup}) are
constructed as tensor products of the supermultiplets in table~\ref{short}, 
and correspondingly 
will be denoted by $(\ell_L^+,\ell_L^-;\ell_R^+,\ell_R^-)_{\rm S}$.
The quantum numbers which denote the 
representations of the $AdS_3$ group $SO(2,2)$
are labeled by numbers $s$ and $\GD$, 
which encode the $AdS$ analogue of spin and mass, respectively. 
They are related to the values of $h_{R}$ and $h_{L}$ 
by $s=h_R-h_L, \GD=h_L+h_R$. 

The massive Kaluza-Klein spectrum of nine-dimensional supergravity on
the $AdS_3\times S^3\times S^3$ background has been computed in
\cite{deBoer:1999rh}. It can be summarized in supermultiplets as
 \begin{eqnarray}\label{tower}
&&   
\bigoplus_{\ell^+\ge0,\ell^-\ge1/2} (\ell^+,\ell^-;\ell^+,\ell^-)_{\rm S}~~\oplus
   \bigoplus_{\ell^+\ge 1/2,\ell^-\ge 0}(\ell^+,\ell^-;\ell^+,\ell^-)_{\rm S} \nonumber\\[1ex]
&&
\qquad\qquad
   \oplus\bigoplus_{\ell^+,\ell^-\ge 0}\big( (\ell^+,\ell^-;\ell^+\!+\ft12,\ell^-\!+\ft12)_{\rm S}
   \oplus (\ell^+\!+\ft12,\ell^-\!+\ft12;\ell^+,\ell^-)_{\rm S} \big).
 \end{eqnarray}
Note that the multiplets $(\ell^+,\ell^-;\ell^+,\ell^-)_{\rm S}$ generically
contain massive fields with spin running from 0 to $\ft32$, whereas
multiplets of the type $(\ell^+,\ell^-;\ell^+\!+\!\ft12,\ell^-\!+\!\ft12)_{\rm S}$
represent massive spin-2 multiplets.

In addition, there is the massless supergravity multiplet
$(\ft12,\ft12;0,0)_{\rm S}\oplus (0,0;\ft12,\ft12)_{\rm S}$,
which consists of the 
vielbein, eight gravitinos, transforming as
\bea
\psi_{\mu}^{I} &:&\;\;
 (\ft12,\ft12;0,0)\oplus (0,0;\ft12,\ft12) \;,
 \label{supercharges}
\eea
under (\ref{so34}),
and topological gauge vectors, corresponding to the 
$SO(4)_{L}\times SO(4)_{R}$ 
gauge group. As a general feature of three-dimensional supergravity
and in accordance with the counting of~table~\ref{short} it
does not contain any physical degrees of freedom.

We should emphasize that except for this supergravity
multiplet and one of the lowest massive spin-$3/2$ multiplets
$(\ft12,\ft12;\ft12,\ft12)_{\rm S}$, all short multiplets
appearing in the Kaluza-Klein spectrum~(\ref{tower})
may combine into long multiplets~(\ref{long})~\cite{deBoer:1999rh}.
The conformal weight of these long representations is not protected by anything
and may vary throughout the moduli space.
This is in contrast to the BPS supergravity spectra usually appearing in 
Kaluza-Klein sphere compactifications.
It gives a distinguished role to the 
supermultiplet~$(\ft12,\ft12;\ft12,\ft12)_{\rm S}$
that we shall analyze in this paper.

\subsection{The lowest supermultiplets and their effective theories}
\label{multiplet}

In this paper, we will construct the three-dimensional
effective actions that describe the couplings of the lowest multiplets
of the full Kaluza-Klein spectrum~(\ref{tower}).
As the $AdS_3\times S^3\times S^3$ background
preserves half of all the supersymmetries, i.e.~16 real supercharges,
the effective three-dimensional  
theory should be (at least) half maximally supersymmetric.
In $D=3$ language this is an ${\cal N}=8$ supergravity.

The lowest supermultiplet in the spectrum is the massless supergravity
multiplet $(0,0;\ft12,\ft12)_{\rm S}\oplus(\ft12,\ft12;0,0)_{\rm S}$. As
noted above, it does not contain propagating degrees of freedom.
It can be effectively described by the difference of two Chern-Simons 
theories~\cite{Achucarro:1987vz,Witten:1988hc} 
\bea
\cal{L} &=& 
{\rm Tr}\,
({\cal A}_{L} \wedge d{\cal A}_{L} + \ft23 {\cal A}_{L}\wedge {\cal A}_{L}\wedge {\cal A}_{L} )
-{\rm Tr}\,
({\cal A}_{R} \wedge d{\cal A}_{R} + \ft23 {\cal A}_{R}\wedge {\cal A}_{R}\wedge {\cal A}_{R}) \;,
\label{CS}
\eea
with connections ${\cal A}_{L}$, ${\cal A}_{R}$ living on the two factors of the 
supergroup~(\ref{supergroup}).

The lowest massive multiplets in the Kaluza-Klein tower (\ref{tower}) are
the degenerate multiplets $(0,\ft12;0,\ft12)_{\rm S}$ and $(0,1;0,1)_{\rm S}$ (together with
$(\ft12,0;\ft12,0)_{\rm S}$ and $(1,0;1,0)_{\rm S}$), to which we will refer as
the spin-$\ft12$ and spin-$1$ multiplet, respectively,
in accordance with their states of maximal spin. 
Their precise representation content is collected in table~\ref{spinspin}.
Coupling of these multiplets requires to extend the topological
Lagrangian~(\ref{CS}) to propagating matter. This is
achieved by three-dimensional gauged 
supergravities~\cite{Nicolai:2000sc,Nicolai:2001ac}.
These theories are obtained as deformations of the ungauged ${\cal N}=8$
and ${\cal N}=16$ theories constructed in~\cite{Marcus:1983hb}
which couple supergravity to scalar fields parametrizing the
coset spaces $SO(8,n)/(SO(8)\times SO(n))$ and $E_{8(8)}/SO(16)$,
respectively.

\begin{table}[bt]
\centering
  \begin{tabular}{c||c|c|} 
   \raisebox{-1.25ex}{$h_L$} \raisebox{1.25ex}{$h_R$} &  
   $\frac{\alpha}{2(1+\alpha)}$ & $\frac{1+2\alpha}{2(1+\alpha)}$ 
   \rule[-2ex]{0pt}{5.5ex}\\
    \hline\hline
   $\frac{\alpha}{2(1+\alpha)}$ & $(0,\ft12;0,\ft12)$ & $(0,\ft12;\ft12,0)$  
   \rule[-1.5ex]{0pt}{4ex}\\ 
    \hline
   $\frac{1+2\alpha}{2(1+\alpha)}$ &  $(\ft12,0;0,\ft12)$ & 
   $(\ft12,0;\ft12,0)$  \rule[-1.5ex]{0pt}{4ex} \\
   \hline
  \multicolumn{3}{c}{\vphantom{I}}\\[.9ex] 
  \end{tabular}
\qquad\;\;
  \begin{tabular}{c||c|c|c|} 
   \raisebox{-1.25ex}{$h_L$} \raisebox{1.25ex}{$h_R$} &  
   $\frac{\alpha}{1+\alpha}$ & $\frac{3\alpha+1}{2(1+\alpha)}$ 
   & $\frac{2\alpha+1}{1+\alpha}$ 
   \rule[-2ex]{0pt}{5.5ex}\\
    \hline\hline
   $\frac{\alpha}{1+\alpha}$ & $(0,1;0,1)$  & $(0,1;\ft12,\ft12)$ & $(0,1;0,0)$ 
   \rule[-1.5ex]{0pt}{4ex}\\ 
    \hline
   $\frac{3\alpha+1}{2(1+\alpha)}$ &  $(\ft12,\ft12;0,1)$ 
   & $(\ft12,\ft12;\ft12,\ft12)$ & 
   $(\ft12,\ft12;0,0)$  \rule[-1.5ex]{0pt}{4ex} \\
     \hline
   $\frac{2\alpha+1}{1+\alpha}$ &  $(0,0;0,1)$ & $(0,0;\ft12,\ft12)$ 
   & $(0,0,0,0)$ \rule[-1.5ex]{0pt}{4ex} \\
  \hline  
  \end{tabular}
      \caption{\small The spin-$\ft12$ multiplet $(0,\ft12;0,\ft12)_{\rm S}$,
      and the massive spin-$1$ multiplet~$(0,1;0,1)_{\rm S}$.} 
\label{spinspin}
\end{table}

Specifically, the scalar sector of these theories is given
by a gauged coset space $\sigma$-model
\bea
\cal{L}_{{\rm matter}} &=& 
e \,{\rm Tr} 
\Big\langle
[{\cal V}^{-1}D_{\mu}{\cal V}]_{\mathfrak{k}}\,
[{\cal V}^{-1}D^{\mu}{\cal V}]_{\mathfrak{k}}
\Big\rangle
+ e \,V(\cal V) +{\rm fermions }
\;
\label{Lag}
\eea
on the target space $G/H=SO(8,n)/(SO(8)\times SO(n))$ and
$G/H=E_{8(8)}/SO(16)$, respectively. Here, the scalar
fields parametrize the $G$-valued matrix ${\cal V}$, 
$[\cdot]_{\mathfrak{k}}$
denotes the projection of the associated Lie algebra $\mathfrak{g}$
onto its noncompact part, and the covariant derivative is given by
\bea
D_{\mu}{\cal V} = \partial_{\mu}{\cal V} + 
g A^{\cM}_{\mu}\,\Theta_{{\cM\cN}}\,t^{\cN}\,{\cal V}\;,
\label{covariant}
\eea
with gauge coupling constant $g$ and 
a constant symmetric matrix $\Theta_{{\cM\cN}}$ (the embedding tensor)
describing the minimal couplings between gauge fields $A^{\cM}_{\mu}$
and symmetry generators $t^{\cN}$ of~$\mathfrak{g}$.
The full effective theory is completely determined by the choice
of this embedding tensor $\Theta_{{\cM\cN}}$.
In particular, the scalar potential $V(\cal V)$ in (\ref{Lag}) is a unique
function of the scalar fields ${\cal V}$ and the embedding 
tensor $\Theta_{{\cM\cN}}$, see~\cite{Nicolai:2000sc,Nicolai:2001ac} 
and (\ref{pot}) below for details.
The topological part~(\ref{CS}) of the Lagrangian 
takes the explicit form
\bea
{\cal L}_{\rm CS} &=& -\ft{1}{4}\,e\,R+
\ft{1}{4}\, g\,\varepsilon^{\mu\nu\rho} \,
A_\mu^{\cM} \,\GTh_{\cM\cN} \Big( \partial_{\nu} A_{\rho}^\cN - 
\ft13 g\, \GTh_{\cK\cS} f^{\cN\cS}{}_{\cL} \, A_{\nu}^\cK A_{\rho}^\cL \Big) 
+{\rm fermions }
 \;,
\label{LCS}
\eea
with structure constants $f^{\cN\cS}{}_{\cL}$ of the algebra $\mathfrak{g}$.
Likewise, the full fermionic Lagrangian is uniquely fixed
in terms of the embedding tensor $\Theta_{{\cM\cN}}$.
The gauge group $G_{0}$ of the effective theory
is identified as the subgroup of $G$ which is
spanned by the generators $\{X_{\cM}=\Theta_{{\cM\cN}}\,t^{\cN}\}$.
At $\Theta_{\cM\cN}=0$, this theory reduces to the ungauged
theory of~\cite{Marcus:1983hb}.
In order to be compatible with supersymmetry, the embedding
tensor needs to satisfy a number of algebraic constraints
which we will describe in more detail below. In turn, 
any solution to these constraints defines a consistent
supersymmetric theory. 

The task of constructing the effective supergravity action
proceeds in several steps. First, one has to determine
the ungauged theory from which the construction starts.
For an ${\cal N}=8$ supersymmetric theory
this is essentially done by comparing 
the bosonic and fermionic degrees of freedom to the 
$16n$ degrees of freedom
described by the theory with scalar target space
$G/H=SO(8,n)/(SO(8)\times SO(n))$. 
More specifically, under $H=SO(8)\times SO(n)$ the physical spectrum of 
this ${\cal N}=8$ supergravity transforms as 
$({\bf 8}_{V}\!\oplus\!{\bf 8}_{C},{\bf n})$,
where ${\bf 8}_{V}$ and ${\bf 8}_C$ denote the vector and conjugate 
spinor representation of $SO(8)$, respectively, and ${\bf n}$ the
vector representation of $SO(n)$. 
For the maximal ${\cal N}=16$ theory 
the spectrum transforms as
${\bf 128}_{S}\oplus {\bf 128}_{C}$ under $H=SO(16)$.
The compact part $G_{\rm c}$ of the desired gauge group $G_{0}$
must then be embedded
into $H$ such that the spectrum decomposes into 
the desired spectrum of the effective theory.
Finally, the corresponding embedding tensor $\Theta_{{\cM\cN}}$
must be determined such that i) it projects the group $G$
onto the desired gauge group $G_{0}$, and ii) it is compatible with
the algebraic constraints imposed by supersymmetry
onto this tensor.

Let us illustrate this with the simplest examples.
The two spin-$\ft12$ multiplets of table~\ref{spinspin}
each contain 16 degrees of freedom.
This suggests that together they are effectively described by 
a gauging of the ${\cal N}=8$ theory
with target space $SO(8,2)/(SO(8)\times SO(2))$.
Indeed, one verifies that the field content of 
$(0,\ft12;0,\ft12)_{\rm S}$ (table~\ref{spinspin}) 
can be lifted from a representation of the gauge group~(\ref{so34})
to an ${\bf 8}_{V}\oplus {\bf 8}_{C}$ of $SO(8)$ with the embedding
\bea
{\bf 8}_{V} \rightarrow (0,\ft12;0,\ft12)\oplus (\ft12,0;\ft12,0) \;,
\qquad
{\bf 8}_{C} \rightarrow (0,\ft12;\ft12,0)\oplus (\ft12,0;0,\ft12) \;,
\label{embedding12}
\eea
while the supercharges~(\ref{supercharges}) lift to the 
spinor representation 
${\bf 8}_{S}$ of $SO(8)$. This corresponds to the
canonical embedding $SO(8)\supset SO(4)\times SO(4)$.
Hence, the two spin-$1/2$ multiplets reproduce the field content
$({\bf 8}_{V}\oplus{\bf 8}_{C},{\bf 2})$ of the ungauged
$SO(8,2)/(SO(8)\times SO(2))$ theory. It remains to
verify that the embedding~(\ref{embedding12})
of the gauge group into $SO(8,2)$ is compatible with the 
constraints imposed by supersymmetry on the embedding tensor 
$\Theta_{{\cM\cN}}$. As it turns out, these requirements 
determine $\Theta_{{\cM\cN}}$ completely up to a free parameter
corresponding to the ratio $\alpha$ of the two sphere radii~\cite{Nicolai:2001ac}.
The effective theory is then completely determined.
Its scalar potential has been further investigated in~\cite{Berg:2001ty}
and indeed reproduces the correct scalar masses
predicted by table~\ref{spinspin}.

The coupling of the spin-$1$ multiplets 
$(0,1;0,1)_{\rm S}\oplus (1,0;1,0)_{\rm S}$
is slightly more involved due to the presence of 
massive vector fields but can be achieved by a
straightforward generalization of the case of 
a single $S^{3}$ compactification~\cite{Nicolai:2003bp,Nicolai:2003ux}.
Here, the effective theory for $128$ degrees of freedom is a gauging
of the ${\cal N}=8$ theory
with coset space $SO(8,8)/(SO(8)\times SO(8))$.
The first thing to verify in this case is 
that the field content of 
$(0,1;0,1)_{\rm S}\oplus (1,0;1,0)_{\rm S}$
(table~\ref{spinspin})
can be lifted from a representation of the gauge group~(\ref{so34})
to an $({\bf 8}_{V}\oplus{\bf 8}_{C},{\bf 8}_{V})$ of $SO(8)\times SO(8)$ 
via the embedding
\bea
&&({\bf 8}_{V},{\bf 1}) \rightarrow (0,\ft12;0,\ft12)\oplus 
(\ft12,0;\ft12,0) \;,
\quad
({\bf 8}_{C},{\bf 1}) \rightarrow (0,\ft12;\ft12,0)\oplus (\ft12,0;0,\ft12) \;,
\nonumber\\[1ex]
&&({\bf 1},{\bf 8}_{V}) \rightarrow (0,\ft12;0,\ft12)\oplus 
(\ft12,0;\ft12,0) \;,
\quad
({\bf 1},{\bf 8}_{C}) \rightarrow (0,\ft12;\ft12,0)\oplus (\ft12,0;0,\ft12) \;.
\label{embedding22}
\eea
This corresponds to the embedding of groups
$SO(8)\times SO(8)\supset SO(8)_{D}\supset SO(4)\times SO(4)$,
where $SO(8)_D$ denotes the diagonal subgroup
of the two $SO(8)$ factors. For instance, (\ref{embedding22})
implies that the bosonic part decomposes as
\bea
({\bf 8}_{V},{\bf 8}_{V}) &\rightarrow&
\Big((0,\ft12;0,\ft12)\oplus (\ft12,0;\ft12,0)\Big)\, \otimes \,
\Big((0,\ft12;0,\ft12)\oplus (\ft12,0;\ft12,0)\Big)
\nonumber\\[1ex]
&=&
(0,1;0,1)\oplus(0,1;0,0)\oplus(0,0;0,1)\oplus
(1,0;1,0)\oplus(1,0;0,0)\oplus(0,0;1,0)
\nonumber\\
&&{}
\oplus 2\cdot(0,0;0,0)\oplus2\cdot
(\ft12,\ft12;\ft12,\ft12) \;,
\nonumber
\eea
in agreement with table~\ref{spinspin}
and its conjugate.
It is important to note that the massive 
spin-$1$ fields 
show up in this decomposition
through their Goldstone scalars.

In order to reproduce the correct coupling for these massive vector fields,
the total gauge group $G_{0}\subset SO(8,8)$ is not just the compact 
factor $G_{\rm c}$~(\ref{so34}), but rather takes the form of a semi-direct product
\bea
G_{0}&=& G_{\rm c}\ltimes T_{12} \;,
\label{GT12}
\eea
with the abelian $12$-dimensional translation group $T_{12}$
transforming in the adjoint representation of 
$G_{\rm c}$~\cite{Nicolai:2003bp}.
In the $AdS_{3}$ vacuum, these translational symmetries are 
broken and the corresponding vector fields gain their masses
in the corresponding Brout-Englert-Higgs mechanism.
The proper embedding of~(\ref{GT12}) into $SO(8,8)$ is again 
uniquely fixed by the constraints imposed by supersymmetry 
on the embedding tensor $\Theta_{{\cM\cN}}$
up to the free parameter $\alpha$~\cite{Nicolai:2003ux}.

Finally, it is straightforward to construct the effective theory
that couples both the spin-$1/2$ and the spin-$1$ supermultiplets
as a gauging of the theory with coset space $SO(8,10)/(SO(8)\times SO(10))$
which obviously embeds the two target spaces described above.

\subsection{The spin-$3/2$ multiplet}
\label{sec:spin32}

The main focus of this paper is the coupling of the
massive spin-$3/2$ multiplet 
$(\ft12,\ft12;\ft12,\ft12)_{\rm S}$
which is contained twice~in (\ref{tower}).
Its $SO(4)\times SO(4)$ representation content
is summarized in table~\ref{spin32}.
As has been discussed in 
the introduction, this 
multiplet is of particular interest
for several reasons. 
In particular, it is the only multiplet
to carry moduli of the theory.
They correspond to
the $SO(4)\times SO(4)$
singlet $(0,0;0,0)$ with conformal dimensions 
$(h_{L},h_{R})=(1,1)$ in table~\ref{spin32}.

\begin{table}[bt]
 \centering
  \begin{tabular}{|c||c|c|c|c|} \hline
   \raisebox{-1.25ex}{$h_L$} \raisebox{1.25ex}{$h_R$} &  
   $\ft{1}{2}$ & $1$ & $\ft32$ & $2$ \rule[-2ex]{0pt}{5.5ex}\\
    \hline\hline
   & & $(\ft12,\ft12;0,0)$ & & \rule[-1.5ex]{0pt}{4ex}\\
   $\ft12$ & $(\ft12,\ft12;\ft12,\ft12)$ & $(\ft12,\ft12;0,1)$ &
   $(\ft12,\ft12;\ft12,\ft12)$ & $(\ft12,\ft12;0,0)$ 
   \rule[-1.5ex]{0pt}{3.5ex}\\
   & & $(\ft12,\ft12;1,0)$ & & \rule[-1.5ex]{0pt}{3.5ex}\\
    \hline
   & &  $(0,0;0,0)$ & &  \rule[-1.5ex]{0pt}{4ex}\\
   & $(1,0;\ft12,\ft12)$ & $(0,1;0,0),(0,0;0,1)$ & $(0,1;\ft12,\ft12)$ & 
     $(0,0;0,0)$ \rule[-1.5ex]{0pt}{3.5ex}\\
   $1$ & $(0,0;\ft12,\ft12)$ & $(1,0;0,0),(0,0;1,0)$ & $(0,0;\ft12,\ft12)$ &
     $(0,1;0,0)$ \rule[-1.5ex]{0pt}{3.5ex}\\
   & $(0,1;\ft12,\ft12)$ & $(0,1;0,1),(1,0;1,0)$ & $(1,0;\ft12,\ft12)$ &
     $(1,0;0,0)$ \rule[-1.5ex]{0pt}{3.5ex}\\
   & &  $(0,1;1,0),(1,0;0,1)$ & & \rule[-1.5ex]{0pt}{3.5ex}\\
    \hline
   & & $(\ft12,\ft12;0,0)$ & & \rule[-1.5ex]{0pt}{4ex}\\
   $\ft32$ & $(\ft12,\ft12;\ft12,\ft12)$ & $(\ft12,\ft12;0,1)$ &
   $(\ft12,\ft12;\ft12,\ft12)$ & $(\ft12,\ft12;0,0)$ 
   \rule[-1.5ex]{0pt}{3.5ex}\\
   & & $(\ft12,\ft12;1,0)$ & & \rule[-1.5ex]{0pt}{3.5ex}\\
    \hline
   & & $(0,0;0,0)$ & & \rule[-1.5ex]{0pt}{4ex}\\
   $2$ & $(0,0;\ft12,\ft12)$ & $(0,0;0,1)$ & $(0,0;\ft12,\ft12)$ 
   & $(0,0;0,0)$ \rule[-1.5ex]{0pt}{3.5ex}\\
   & & $(0,0;1,0)$ & & \rule[-1.5ex]{0pt}{3.5ex}\\
    \hline
  \end{tabular}
 \caption{\small The massive spin-$3/2$ supermultiplet 
 $(\ft12,\ft12;\ft12,\ft12)_{\rm S}$.}
\label{spin32}
\end{table}

In this paper we will construct the coupling of this multiplet
to the massless supergravity multiplet.
In analogy to the aforementioned couplings of the spin-$\ft12$ and 
spin-$1$ multiplet to ${\cal N}=8$ supergravity,
a natural candidate for the effective theory
might be an ${\cal N}=8$ gauging of the effective theory
with coset space $SO(8,16)/(SO(8)\times SO(16))$,
reproducing the correct number of 128 bosonic degrees of freedom.
(The appearance of massive spin-$\ft32$
fields would then require some analogue of the 
dualization taking place in the scalar/vector sector.)
Let us check the representation content of table~\ref{spin32}.
It is straightforward to verify that
the states of this multiplet may be lifted from a representation
of the gauge group~(\ref{so34}) to a representation 
$({\bf 8}_{V}\oplus {\bf 8}_{C},{\bf 8}_{V}\oplus {\bf 8}_{C})$
of an $SO(8)_{L}\times SO(8)_{R}$ according to 
\bea
&&({\bf 8}_{V},{\bf 1}) \rightarrow (0,0;0,0)\oplus 
(0,0;0,0)\oplus (1,0;0,0)\oplus (0,1;0,0)\;,
\nonumber\\
&&({\bf 8}_{C},{\bf 1}) \rightarrow (\ft12,\ft12;0,0)\oplus 
(\ft12,\ft12;0,0) \;,
\qquad
({\bf 8}_{S},{\bf 1}) \rightarrow (\ft12,\ft12;0,0)\oplus (\ft12,\ft12;0,0) \;,
\nonumber\\
&&({\bf 1},{\bf 8}_{V}) \rightarrow(0,0;0,0)\oplus (0,0;0,0)\oplus 
(0,0;1,0)\oplus (0,0;0,1) \;,
\quad
\nonumber\\
&&({\bf 1},{\bf 8}_{C}) \rightarrow (0,0;\ft12,\ft12)\oplus 
(0,0;\ft12,\ft12)\;,
 \qquad
({\bf 1},{\bf 8}_{S}) \rightarrow (0,0;\ft12,\ft12)\oplus (0,0;\ft12,\ft12) \;.
\label{embedding33}
\eea
This corresponds to an embedding of groups according to
\bea\label{so4embed}
SO(4)_{L} = {\rm diag}[SO(4)\times SO(4)]\subset SO(4)\times SO(4) \subset SO(8)_{L}\;,
\eea 
and similarly for $SO(4)_{R}$.
In order to be described as a gauging of the ${\cal N}=8$
theory, the field content would have to be further lifted 
to the $({\bf 8}_{V}\oplus {\bf 8}_{C},{\bf 16})$
of $SO(8)\times SO(16)$.
This is only possible, if $SO(8)_{R}$ is entirely embedded
into the $SO(16)$. In turn, this implies that the gravitino
fields transforming in the $({\bf 8}_{S},{\bf 1})$
of the ${\cal N}=8$ theory would decompose
as $(\ft12,\ft12;0,0)\oplus (\ft12,\ft12;0,0)$, in contrast to the 
supercharges~(\ref{supercharges})
of the Kaluza-Klein spectrum.
We conclude that the massive spin-$3/2$ multiplet
cannot be described as a gauging of the $SO(8,16)/(SO(8)\times SO(16))$
theory.

Rather we will find that the effective theory describing 
this multiplet is a maximally supersymmetric
gauging of the ${\cal N}=16$ theory in its broken phase.
Half of the supersymmetry is broken down to ${\cal N}=8$ and
correspondingly eight gravitinos acquire mass via a super-Higgs 
mechanism.
As a first check
we observe that the total number of degrees of freedom
collected in table~\ref{spin32} indeed equals the $16^2=256$
of the maximal theory. 
More specifically, according to~(\ref{embedding33}),
the total spectrum can be lifted to an
$({\bf 8}_{V}\oplus {\bf 8}_{S},{\bf 8}_{V}\oplus {\bf 8}_{C})$
of $SO(8)_{L}\times SO(8)_{R}$ 
and thus further to $SO(16)$ 
according to
 \bea
 \label{so16deco} 
&&   \f{16}\rightarrow (\f{8}_C,\f{1})\oplus (\f{1},\f{8}_S)\;, \nonumber\\
&&   \f{128}_S\rightarrow (\f{8}_V,\f{8}_V)\oplus (\f{8}_S,\f{8}_C)\;, \quad
   \f{128}_C\rightarrow (\f{8}_S,\f{8}_V)\oplus (\f{8}_V,\f{8}_C)\;.
 \eea
This corresponds to the
canonical embedding $SO(16)\supset SO(8)_{L}\times SO(8)_{R}$
and an additional triality rotation.
Moreover, this lifts the spectrum
precisely to the ${\bf 128}_{S}\oplus {\bf 128}_{C}$
field content of the maximal ${\cal N}=16$
theory with scalar target space $G/H=E_{8(8)}/SO(16)$.
In the next section we shall describe this embedding 
in full detail.

\section{The ${\cal N}=16$ Supergravity Action}\label{3action} 

In this section, we construct the effective three-dimensional
action that describes the coupling of the massive spin-$3/2$ multiplet
$(\ft12,\ft12;\ft12,\ft12)_{{\rm S}}$.
We proceed in three steps. The general maximal 
(${\cal N}=16$) supergravity in three dimensions is reviewed in section 3.1.
It is given as a gauging of the maximal 
theory with target space $E_{8(8)}/SO(16)$
which is entirely specified by the choice of the constant 
embedding tensor.
In section 3.2~we determine the total gauge group $G_{0}$
of the effective theory taking into account the
extra translational directions related to the 
presence of massive vector fields. We further describe
the embedding of $G_{0}$ into the global symmetry 
group~$E_{8(8)}$.
In section 3.3 we determine the embedding tensor
$\Theta_{\cal{M}\cal{N}}$ related to this gauge group, and show
that it is fixed up to four free parameters
by the algebraic consistency constraints
imposed by supersymmetry.
Together with the general results of section 3.1
this completely determines the effective theory.
We show in section 3.4 that the requirement
of the existence of an $AdS$ ground state
with the correct isometry group~(\ref{supergroup})
further fixes two of the parameters, such that
the final theory has only two free parameters
corresponding to the radii of the two $S^{3}$ spheres.

\subsection{Maximal supergravity in $D=3$}\label{gauging}

We start with a brief summary of the important facts about
${\cal N}=16$ supergravity in three dimensions
following~\cite{Nicolai:2000sc}.
The spectrum consists of the supergravity multiplet which contains 
the vielbein $e_{\mu}^a$ and $16$ Majorana gravitino
fields $\psi_{\mu}^I$, $I=1,...,16$. 
The propagating matter consists of 128 real scalar fields and 
128 Majorana fermions $\chi^{\dot{A}}$, $\dot{A}=1,...,128$. 
They combine into an irreducible supermultiplet 
for the maximal amount of $32$ real supercharges. 
The scalar fields parametrize the coset space 
$E_{8(8)}/SO(16)$. 
Equivalently they can be represented by an $E_{8(8)}$ valued matrix
$\mathcal{V}(x)$, which transforms under global $E_{8(8)}$ and 
local $SO(16)$ transformations as follows:
 \bea\label{coset}
  \mathcal{V}(x)\rightarrow g\mathcal{V}(x)h^{-1}(x), \qquad
  g\in E_{8(8)}\;,\quad h(x)\in SO(16)\;.
 \eea  
Specifying~(\ref{Lag}) to the ${\cal N}=16$ case,
the maximal supergravity Lagrangian 
up to quartic couplings in the fermions is given by
 \bea
 \label{action}
  \mathcal{L}&=&-\frac{1}{4}eR+\frac{1}{4}e\,{\cal P}^{\mu A}{\cal P}_{\mu}^A+
  \frac{1}{2}\varepsilon^{\mu\nu\rho}\overline{\psi}_{\mu}^I {\cal D}_{\nu}\psi_{\rho}^I-
  \frac{i}{2}e\,\bar{\chi}^{\dot{A}}\gamma^{\mu}{\cal D}_{\mu} \chi^{\dot{A}}
  -\ft12 e \bar{\chi}^{\dot{A}}\gamma^{\mu}\gamma^{\nu}\psi_{\mu}^I
  \Gamma^{I}_{A\dot{A}}\,{\cal P}^{A}_{\nu}
  \nonumber\\[1ex]
  &&{}
+\frac{1}{2}eg\,A_1^{IJ}\,\bar{\psi_{\mu}}^I
  \gamma^{\mu\nu}\psi_{\nu}^J + ieg\,A_2^{I\dot{A}}\,\bar{\chi}^{\dot{A}}
  \gamma^{\mu}\psi_{\mu}^I 
  + \frac{1}{2}eg\,A_3^{\dot{A}\dot{B}}\, \bar{\chi}^{\dot{A}}\chi^{\dot{B}}
    \nonumber\\[1ex]
  &&
{}
  +
\ft{1}{4}\, g\,\varepsilon^{\mu\nu\rho} \,
A_\mu^{\cal{M}} \,\GTh_{\cal{M}\cal{N}} \Big( \partial_{\nu} A_{\rho}^{\cal{N}} - \ft13 g\, \GTh_{\cal{K}\cal{L}} f^{\cal{N}\cal{L}}{}_{\cal{P}} \, 
A_{\nu}^{\cal{K}} A_{\rho}^{\cal{P}} \Big) 
-e\,V(\cal V)
\;.
\eea 
As discussed in~(\ref{covariant}) above, the theory is entirely encoded
in the symmetric constant matrix $\Theta_{{\cal{M}\cal{N}}}$.
It describes the minimal coupling of vector fields to scalars
according to
 \bea\label{current}
  \mathcal{V}^{-1}D_{\mu}\mathcal{V}~\equiv~
  \mathcal{V}^{-1} \partial_{\mu}{\cal V} + 
g A^{\cal{M}}_{\mu}\,\Theta_{{\cal{M}\cal{N}}}\,
(\mathcal{V}^{-1} t^{\cal{N}}\,{\cal V})
~\equiv~
\ft{1}{2}{\cal Q}_{\mu}^{IJ}X^{IJ}
   +{\cal P}_{\mu}^AY^A\;,
 \eea
with $X^{IJ}$ and $Y^{A}$ labeling the 120 compact and 128 noncompact
generators of $E_{8(8)}$, respectively.\footnote{See appendix \ref{A1} for 
our $E_{8(8)}$ and $SO(16)$ conventions.}
The fermionic mass terms in~(\ref{action}) are defined as linear functions
of $\Theta_{\cal{M}\cal{N}}$ via
 \bea\label{A123}
   A_1^{IJ}&=&\frac{8}{7} \theta \delta_{IJ} +\frac{1}{7}T_{IK|JK}\;, 
   \quad
   A_2^{I\dot{A}}=-\frac{1}{7}\Gamma_{A\dot{A}}^JT_{IJ|A}\;, \nonumber\\
   A_3^{\dot{A}\dot{B}}&=&2\theta\delta_{\dot{A}\dot{B}}+\frac{1}{48}
   \Gamma_{\dot{A}\dot{B}}^{IJKL}T_{IJ|KL},
\eea
with $SO(16)$ gamma matrices $\Gamma^I_{A\dot{B}}$, 
and the $T$-tensor 
\bea\label{T}
  T_{\cal M|\cal N}  &=&\mathcal{V}^{\cal K}{}_{\cal M}
  \mathcal{V}^{\cal L}{}_{\cal N}
  \Theta_{\cal K\cal L}\;.
\eea
The scalar potential $V(\cal V)$ is given by
\bea
\label{pot}
  V=-\frac{g^2}{8}\,\big(A_1^{KL}A_1^{KL}-\frac{1}{2}A_2^{K\dot{A}}
  A_2^{K\dot{A}}\big)\;.
\eea
For later use we note the condition of stationarity of this potential:
 \bea\label{stationary}
\delta\, V =0 \qquad
\Longleftrightarrow
\qquad  
3 A_1^{IM}A_2^{M\dot{A}}=A_3^{\dot{A}\dot{B}}A_2^{I\dot{B}}\;.
 \eea
The quartic fermionic couplings and the supersymmetry transformations
of~(\ref{action}) can be found in~\cite{Nicolai:2000sc}. 
For consistency of the theory, the embedding tensor $\Theta_{\cal M\cal N}$
needs to satisfy two algebraic constraints.
First, $\Theta$ as an element of the symmetric $E_{8(8)}$ tensor product
\bea
  (\f{248}\otimes \f{248})_{\text{sym}}&=&
  \f{1}\oplus \f{3875} \oplus \f{27000}
 \;,
\eea
is required to only live in the ${\bf 1}\oplus{\bf 3875}$ representation,
i.e.~to satisfy the projection constraint
$(\mathbb{P}_{\f{27000}})\,\Theta\equiv0\,$.
Explicitly, this constraint takes the form~\cite{Koepsell:1999uj}
\bea
\Theta_{\cal M\cal N}
+\frac{1}{62}\,\eta_{\cal M\cal N}\,\eta^{\cal K\cal L}\,
\Theta_{\cal K\cal L}
+\frac{1}{12}\, \eta_{\cal P\cal Q} f^{\cal K\cal P}{}_{\cal M} 
f^{\cal L\cal Q}{}_{\cal N}\,\Theta_{\cal K\cal L}&=& 0 \;,
\label{conex}
\eea
with the $E_{8(8)}$ structure constants $f^{\cal M\cal N}{}_{\cal K}$ 
and Cartan-Killing form $\eta^{\cal M\cal N}=\frac{1}{60}\,
  f^{\cM\cK}_{\hspace{1.5em}\cL}\,f^{\cN\cL}_{\hspace{1.5em}\cK}$.
Second, closure of the gauge group requires the $E_{8(8)}$ covariant
quadratic condition
\bea\label{quad}
  \Theta_{\cal K\cal P}\Theta_{{\cal L}(\cal M}\,f^{\cal K\cal L}
  {}_{\cal N)}=0\;,
\eea
on $\Theta\,$. In turn, any solution of (\ref{conex}) and (\ref{quad})
defines a consistent maximally supersymmetric theory~(\ref{action})
in three dimensions.

It should be mentioned that although the formulation~(\ref{action})
seems to restrict vector couplings to the topological Chern-Simons term,
any supergravity theory, including those with propagating Yang-Mills
kinetic term can be casted into this form upon proper enhancement
of the gauge group $G_{0}$~\cite{Nicolai:2003bp}.
We will discuss this in more detail in the next section.
In the following we will identify the particular embedding tensor
$\Theta_{\cM\cN}$ that corresponds to
the effective theory coupling the massive spin-$3/2$ multiplet 
on the $AdS_3\times S^3\times S^3$ background.

\subsection{The gauge group $G_{0}$}
\label{embed}

In this section, we will identify the full gauge group
of the effective three-dimensional theory and determine
its embedding into the global $E_{8(8)}$ symmetry group
of the ungauged theory.
To this end we need to first review the general structure
of gauge groups in three dimensions.
It has been shown in~\cite{Nicolai:2003bp}
that generically the gauge group of a
three-dimensional Chern-Simons gauged
supergravity~(\ref{action}) is of the form
\bea
G_{0}&=&
{G}_{\rm c} \ltimes (\hat{T}_p, {T}_\nu)
\;.
\label{gengau}
\eea
In our case, ${G}_{\rm c}$ denotes the compact
gauge group~(\ref{so34}) which from the Kaluza-Klein
origin of the theory is 
expected to be realized by propagating vector fields.
In the Chern-Simons formulation given above,
this compact factor needs to be amended by the 
nilpotent translation group 
${T}_\nu$ whose $\nu={\rm dim}\,{G}_{\rm c}$
generators transform in the adjoint representation
of ${G}_{\rm c}$. This allows an alternative formulation
of the theory~(\ref{action}) in which 
part of the scalar sector is redualized 
into propagating vector fields gauging the group ${G}_{\rm c}$
which accordingly appear with a conventional 
Yang-Mills term.
In other words, a Yang-Mills gauged theory with 
gauge group ${G}_{\rm c}$
is on-shell equivalent to a Chern-Simons gauged
supergravity theory with gauge group 
$G_{\rm c}\ltimes \mathcal{T}_{\nu}$.
The third factor $\hat{T}_p$ in~(\ref{gengau})
is spanned by $p$ nilpotent translations 
transforming in some representation of ${G}_{\rm c}$ and
closing into ${T}_\nu$.
This part of the gauge group is completely broken
in the vacuum and gives rise to $p$ massive vector fields.
Specifically, the algebra underlying~(\ref{gengau}) reads
 \bea\label{semidirect}
&&  
[\mathcal{J}^m,\mathcal{J}^n]=f^{mn}_{\hspace{1.2em}k}\,
\mathcal{J}^k\;,
\qquad 
  [\mathcal{J}^m,\mathcal{T}^{\underline{n}}]
  =f^{mn}_{\hspace{1.2em}k}\,\mathcal{T}^{\underline{k}}\;, 
\qquad
  [\mathcal{T}^{\underline{m}},\mathcal{T}^{\underline{n}}]=0\;,
\nonumber\\[1ex] 
&&
  [\mathcal{J}^m,\hat{\mathcal{T}}^{\alpha}]
  =t^{m\alpha}_{\hspace{1.0em}\beta}\,\hat{\mathcal{T}}^{\beta}\;, 
  \qquad
  [\hat{\mathcal{T}}^{\alpha},\hat{\mathcal{T}}^{\beta}]
  =t^{\alpha\beta}_{\hspace{1.0em}m}\,\mathcal{T}^{\underline{m}}\;, 
  \qquad
  [\mathcal{T}^{\underline{m}},\hat{\mathcal{T}}^{\alpha}]=0\;,
 \eea
with $\mathcal{J}^m$, $\mathcal{T}^{\underline{n}}$, and $\hat{\mathcal{T}}^{\alpha}$
generating ${G}_{\rm c}$, ${T}_\nu$, and $\hat{T}_p$, respectively. 
The $f^{mn}_{\hspace{1.2em}k}$ are the structure constants of ${G}_{\rm c}$
while the $t^{m\alpha}_{\hspace{1.0em}\beta}$ denote the 
representation matrices for the $\hat{\mathcal{T}}^{\alpha}$.
Indices $m, n, \dots$ are raised/lowered with the Cartan-Killing form of ${G}_{\rm c}$;
raising/lowering of 
indices $\alpha$, $\beta$ requires a symmetric ${G}_{\rm c}$
invariant tensor $\kappa^{\alpha\beta}$.

To begin with, we have to reconcile the structure~(\ref{gengau})
with the spectrum collected in table~\ref{spin32}.
With ${G}_{\rm c}=SO(4)_{L}\times SO(4)_{R}$ from~(\ref{so34}),
${T}_\nu$ transforms in the adjoint representation
$(1,0;0,0)\oplus (0,1;0,0)\oplus(0,0;1,0)\oplus (0,0;0,1)$.
table~\ref{spin32} exhibits $34$ additional massive vector
fields, transforming in the 
$2\!\cdot\!(\frac12,\frac12;\frac12,\frac12)\oplus2\!\cdot\!(0,0;0,0)$ of 
${G}_{\rm c}$.
In total, we thus expect a gauge group
$G_{0}={G}_{\rm c} \ltimes (\hat{T}_{34}, {T}_{12})$
of dimension ${\rm dim}\,G_{0}=12+12+34=58$.
Next, we have to identify this group within $E_{8(8)}$.
To this end, it proves useful to first consider the embedding
of ${G}_{\rm c}$ into $E_{8(8)}$ according to the decompositions
\bea
E_{8(8)} ~\supset~
\left\{
\begin{array}{lcr}
\supset & SO(16)& \supset  \\[2ex]
\supset & SO(8,8)& \supset
\end{array}
\right\}
~\supset~
SO(8)_{L} \times SO(8)_{R} 
~\supset~ SO(4)_{L} \times SO(4)_{R} \;,
\label{embbb}
\eea
with the two embeddings of $SO(8)_{L} \times SO(8)_{R}$ given by
 \bea
SO(16)  &:& \quad  \f{16}\rightarrow (\f{8}_C,\f{1})\oplus (\f{1},\f{8}_S)\;, \quad
 \f{128}_S\rightarrow (\f{8}_V,\f{8}_V)\oplus (\f{8}_S,\f{8}_C)\;, 
 \nonumber\\[2ex]
 SO(8,8) &:& \quad  \f{16}\rightarrow (\f{8}_V,\f{1})\oplus (\f{1},\f{8}_V)\;, \quad
 \f{128}_S\rightarrow (\f{8}_C,\f{8}_S)\oplus (\f{8}_S,\f{8}_C)\;.
 \label{16vs88} 
  \eea
Accordingly, the group $E_{8(8)}$ decomposes as
\bea
\f{248} ~\rightarrow ~
\Big\{
(\f{28},\f1) \oplus (\f1,\f{28}) \oplus
 (\f{8}_C,\f{8}_S)\Big\} 
 ~\oplus~
 \Big\{ 
  (\f{8}_V,\f{8}_V)\oplus (\f{8}_S,\f{8}_C) \Big\}
  \;,
\eea
and further according to~(\ref{embedding33}). 
Here curly brackets indicate the splitting into
its compact and noncompact part and $\f{28}=\f{8}\wedge\f{8}$.
We have already discussed in section~\ref{sec:spin32}
that with this embedding
the noncompact part of $E_{8(8)}$ precisely
reproduces the bosonic 
spectrum of table~\ref{spin32}.

In order to identify the embedding of the full gauge group
$G_{0}={G}_{\rm c} \ltimes (\hat{T}_{34}, {T}_{12})$
we further consider the decomposition
of $E_{8(8)}$ according to
\bea
E_{8(8)} \supset SO(8,8) \supset 
SO(6,6)\times SO(2,2)
\supset 
SO(6,6)\times SO(1,1)\times SO(1,1)
\;,
\eea
and its grading  w.r.t.~these two $SO(1,1)$ factors
which is explicitly given in table~\ref{grading}.
From this table we can infer that properly identifying
\bea\label{e8grade}
{G}_{\rm c}~\subset~ \f{66}_0^0\;,\qquad
{T}_{12} ~=~ \f{12}_{0}^{+1} \;,
\qquad
\hat{T}_{34} ~\subset~ 
\f{32}_{-1/2}^{+1/2} ~\oplus~ \overline{\f{32}}_{+1/2}^{+1/2}
~\oplus~ \f{1}^{+1}_{-1} ~\oplus~ \f{1}^{+1}_{+1}
\;,
\label{GTT}
\eea
precisely reproduces the desired algebra structure~(\ref{semidirect}).
We have thus succeeded in identifying the algebra~$\mathfrak{g}_{0}$ 
underlying the full gauge group
$G_{0}={G}_{\rm c} \ltimes (\hat{T}_{34}, {T}_{12})$,
which is entirely embedded in the ``upper light cone''
of table~\ref{grading}.
In the next section, we will explicitly construct
the embedding tensor $\Theta_{\cM\cN}$
projecting onto this algebra, and show
that it is indeed compatible with the algebraic
constraints~(\ref{conex}), (\ref{quad}) 
imposed by supersymmetry.

 \begin{table}[bt]
  \begin{center}
   $\begin{array}{cccccccccccc} \rule[-0.4cm]{0cm}{0.4cm}
     \f{1}^{+1}_{-1} &&    \f{12}_{0}^{+1} &&  \f{1}^{+1}_{+1}
     \\ \rule[-0.4cm]{0cm}{0.4cm}
     & \f{32}_{-1/2}^{+1/2} & & \overline{\f{32}}_{+1/2}^{+1/2}
       \\ \rule[-0.4cm]{0cm}{0.4cm}
     \f{12}_{-1}^0 & & \f{66}_0^0 + \f{1}_0^0 + \f{1}_0^0 &&
     \f{12}_{+1}^{0}
   \\ \rule[-0.4cm]{0cm}{0.4cm}
     & \overline{\f{32}}_{-1/2}^{-1/2} & & {\f{32}}_{+1/2}^{-1/2}
   \\ \rule[-0.4cm]{0cm}{0.4cm}
      \f{1}^{-1}_{-1} &&    \f{12}_{0}^{-1} &&  \f{1}^{-1}_{+1}
    \end{array}$
  \end{center}
  \caption{\small Grading of $E_{8(8)}$ according to
$SO(6,6)\times SO(1,1)\times SO(1,1)$.
For later reference, we denote by $SO(1,1)_{a}$ the factor
responsible for the grading from left to right and by 
$SO(1,1)_{b}$ the factor
responsible for the grading from top to bottom.}
\label{grading}
 \end{table}

\subsection{The embedding tensor}
\label{constr}

In this section, we will explicitly construct
the embedding tensor $\Theta_{\cM\cN}$
projecting onto the Lie algebra $\mathfrak{g}_{0}$
of the desired gauge group
$G_{0}={G}_{\rm c} \ltimes (\hat{T}_{34}, {T}_{12})$
identified in the previous section.
The embedding tensor then uniquely 
defines the effective action~(\ref{action}).
We start from the $SO(4)\times SO(4)$ basis
of $E_{8(8)}$ defined in appendix~\ref{A3}.
In this basis, the grading of table~\ref{grading} 
refers to the charges of the 
generators $X^{0\hat{0}}$ and $X^{\bar0\hat{\bar0}}$.
We further denote the generators of
$G_{c}$ and ${T}_{12}$ within table~\ref{grading} as
 \begin{equation}
   \f{66}^0_0~\supset~ \f{3}_L^{+}\oplus
   \f{3}_L^{-}\oplus\f{3}_R^{+}\oplus\f{3}_R^{-}\;, 
   \qquad \mbox{and}\quad
   \f{12}_{0}^{+1}~=~\f{\hat3}{}_L^{+}\oplus
   \f{\hat3}{}_L^{-}\oplus\f{\hat3}{}_R^{+}\oplus\f{\hat3}{}_R^{-}\;,
 \label{g1}
 \end{equation}
respectively, with the labels $L$, $R$, $\pm$ referring to the four 
factors of~(\ref{so34}), i.e.\
$\f{3}_L^{+}=(1,0;0,0)$, $\f{3}_L^{-}=(0,1;0,0)$, etc.
Similarly, we will identify the generators of $\hat T_{34}$ among 
 \begin{equation}
   \qquad \f{32}^{+1/2}_{-1/2}\equiv\f{16}_{-}^{(1)} \oplus \f{16}_{-}^{(2)}\;,
   \qquad
   \overline{\f{32}}^{+1/2}_{+1/2} \equiv{\f{16}}_{+}^{(1)}
   \oplus {\f{16}}_{+}^{(2)}\;, 
   \qquad
   \f{1}^{+1}_{-1} \equiv \f{1}_{-}\;,\quad
   \f{1}^{+1}_{+1} \equiv \f{1}_{+}\;,
   \label{g2}
 \end{equation}
where $\bf{16}$ denotes the
$(\ft12,\ft12;\ft12,\ft12)$ of $SO(4)\times SO(4)$,
and we use subscripts $(1), (2)$, $\pm$ in order to distinguish
the identical representations.
The split of the $\bf{32}$ representations into two copies of 
the $\bf{16}$ is chosen such that the algebra 
closes according to
 \bea
   [\f{16}_+^{(1)},\f{16}_-^{(1)}]&\subset& \hat{\f{3}}_L^- \oplus \hat{\f{3}}_R^-\;, 
   \qquad
     [\f{16}_+^{(2)},\f{16}_-^{(2)}]~\subset~ \hat{\f{3}}_L^- \oplus \hat{\f{3}}_R^-\;,
    \nonumber\\[1ex]
{}  [\f{16}_+^{(1)},\f{16}_-^{(2)}]&\subset& \hat{\f{3}}_L^+ \oplus \hat{\f{3}}_R^+\;,
   \qquad
  [\f{16}_+^{(2)},\f{16}_-^{(1)}]~\subset~ \hat{\f{3}}_L^+ \oplus \hat{\f{3}}_R^+\;.
 \eea

The embedding tensor $\Theta_{\cM\cN}$ is an object in the
symmetric tensor product of two adjoint representations
of $E_{8(8)}$. 
It projects onto the Lie algebra of the gauge group
according to 
$\mathfrak{g}_{0}=\langle X_{\cM}\equiv\Theta_{\cM\cN}\,t^{\cN}\rangle$.
We start from the most general ansatz for $\Theta_{\cM\cN}$
that has entries only on the generators (\ref{g1}), (\ref{g2}).
Since the $\Theta_{\cM\cN}$ relevant for our theory moreover
is an $SO(4)\times SO(4)$ invariant tensor,
it can only have non-vanishing entries contracting coinciding
representations, e.g. $\Theta_{\f3_L^+,\f{\hat3}_L^+}$,
$\Theta_{{\f{16}}_{+}^{(1)},\f{16}^{(2)}_-}$, etc.
Using computer algebra (Mathematica), we can
then implement the algebraic constraint~(\ref{conex}).
As one of the main results of this paper,
we find that this constraint determines the embedding
tensor $\Theta$ with these properties up to five free constants 
$\gamma$, $\beta_{1}$, $\beta_{2}$, $\beta_{3}$, $\beta_{4}$,
in terms of which it takes the 
form\footnote{Here we have used a somewhat symbolic notation for $\Theta$,
indicating just the multiples of the identity matrix that
$\Theta$ takes in the various blocks.}
 \bea
 \label{theta2}
&&
    \Theta_{\f3_L^+,\f{\hat3}{}_L^+} =\beta_{1}\;, \quad
    \Theta_{\f3_L^-,\f{\hat3}{}_L^-}=\beta_{2}\;, \quad
    \Theta_{\f3_R^+,\f{\hat3}{}_R^+}=\beta_{3}\;, \quad
    \Theta_{\f3_R^-,\f{\hat3}{}_R^-}=\beta_{4}\;, \nonumber\\[2ex]
&&
   \Theta_{\f{1}_+,\f{1}_-}=
    \Theta_{\f{\hat3}{}_R^+,\f{\hat3}{}_R^+}=
    \Theta_{\f{\hat3}{}_R^-,\f{\hat3}{}_R^-}=
    -\Theta_{\f{\hat3}{}_L^+,\f{\hat3}{}_L^+}=
    -\Theta_{\f{\hat3}{}_L^-,\f{\hat3}{}_L^-}=\gamma \;, 
    \nonumber\\[2ex]
&&
   \Theta_{{\f{16}}_{+}^{(1)},\f{16}^{(1)}_-}
    ~=~-\frac{1}{32\sqrt{2}}(\beta_2+\beta_4)\;, \qquad
    \Theta_{{\f{16}}_{+}^{(1)}, 
    \f{16}^{(2)}_-}~=~-\frac{1}{32\sqrt{2}}(\beta_1
    +\beta_3)\;, 
\nonumber\\[1ex]
&&
    \Theta_{{\f{16}}^{(2)}_{+},
    \f{16}^{(1)}_-}~=~\frac{1}{32\sqrt{2}}(\beta_1
    -\beta_3)\;, \qquad
    \Theta_{{\f{16}}^{(2)}_{+},\f{16}^{(2)}_-}
    ~=~-\frac{1}{32\sqrt{2}}(\beta_2-\beta_4)\;.
 \eea
A priori, it seems quite surprising that the 
constraint~(\ref{conex}) still leaves five free constants in $\Theta$ ---
the ${\bf 27000}$ representation of $E_{8(8)}$ gives
rise to $1552$ different $SO(4)\times SO(4)$ representations
that are separately imposed as constraints on 
our general ansatz for $\Theta$.

In order to satisfy the full set of consistency constraints it remains
to impose the quadratic constraint~(\ref{quad}) on the embedding
tensor~$\Theta_{\cM\cN}$. Again using computer algebra, we can
compute the form of this constraint for the embedding tensor~(\ref{theta2})
and find that it reduces to a single condition on the parameters:
\bea\label{quadr}
  \beta_1^2+\beta_2^2 &=& \beta_3^2 + \beta_4^2\;.
\eea 
This suggests the parametrization as
\bea
\beta_{1} = \kappa\sin\alpha_{1}\;,\quad
\beta_{2} = \kappa\cos\alpha_{1}\;,\quad
\beta_{3} = \kappa\sin\alpha_{2}\;,\quad
\beta_{4} = \kappa\cos\alpha_{2}\;.
\label{alpha12}
\eea

Altogether we have shown, that 
there is a four parameter family of maximally supersymmetric
theories, described by the embedding tensor~(\ref{theta2}), which satisfies
all the consistency constraints~(\ref{conex}),~(\ref{quad}).

For generic values of the parameters, one verifies that the rank of
the induced gauge group is indeed $58$ as expected.\footnote{Let 
us note that the degenerate case $\kappa=0$
induces a theory with 
14-dimensional nilpotent abelian gauge group, as can be seen from 
(\ref{theta2}). This particular gauge group had already been identified 
in~\cite{Fischbacher:2003yw}.}
In particular, (\ref{theta2}), (\ref{quadr}) imply 
that on the block of $\f{16}$ representations
one finds
\bea
\Theta_{\cM\cN}\:  t^{\cM}\otimes t^{\cN}\Big|_{\f{16}} &=&
-\frac\kappa{16\sqrt{2}}\,
\Big(
\f{16}_{+}^{(1)}\cos(\ft12(\alpha_{1}\!-\!\alpha_{2}))-
\f{16}_{+}^{(2)}\sin(\ft12(\alpha_{1}\!-\!\alpha_{2}))
\Big)\otimes
\nonumber \\
&&{}
\Big(
 \f{16}_{-}^{(1)}\cos(\ft12(\alpha_{1}\!+\!\alpha_{2}))-
 \f{16}_{-}^{(2)}\sin(\ft12(\alpha_{1}\!+\!\alpha_{2}))
\Big)
\;.
\eea
This implies that out of the 64 generators $\f{16}_{\pm}^{(1)}$, $\f{16}_{\pm}^{(2)}$,
only the 32 combinations
\bea
\f{16}_{+}&\equiv&
\f{16}_{+}^{(1)}\cos(\ft12(\alpha_{1}\!-\!\alpha_{2}))-
\f{16}_{+}^{(2)}\sin(\ft12(\alpha_{1}\!-\!\alpha_{2}))
\;,
\nonumber\\
\f{16}_{-}&\equiv&
\f{16}_{-}^{(1)}\!\cos(\ft12(\alpha_{1}\!+\!\alpha_{2}))-
\!\f{16}_{-}^{(2)}\sin(\ft12(\alpha_{1}\!+\!\alpha_{2}))
\;,
\eea
form part of the gauge group.
These correspond to the $2\!\cdot\!(\ft12,\ft12;\ft12,\ft12)$ generators in $\hat T_{34}$.
The complete gauge algebra spanned by the generators
$X_{\cM}\equiv \Theta_{\cM\cN}\:  t^{\cN}$ is precisely 
of the form anticipated in~(\ref{semidirect}).

Let us stress another important property of the embedding 
tensor~(\ref{theta2}):
it is a singlet not only under the $SO(4)\times SO(4)$, but also under the 
$SO(1,1)_{a}$
generating the grading from left to right in table~\ref{grading}.
In other words, the resulting $\Theta$ 
contracts only representations for which 
these particular charges add up to zero.
As a consequence the gauged supergravity~(\ref{action}) 
in addition to the local gauge symmetry 
$G_{0}={G}_{\rm c} \ltimes (\hat{T}_{34}, {T}_{12})$  
is invariant under the action of the global symmetry $SO(1,1)_{a}$.
We will discuss the physical consequences of this extra symmetry 
in section~\ref{sec:pot} below.

\subsection{Ground state and isometries}
\label{ground}

In the previous section we have found a four-parameter family 
of solutions $\Theta_{\cM\cN}$~(\ref{theta2}) to the algebraic 
constraints~(\ref{conex}),~(\ref{quad})
compatible with the gauge 
algebra~$G_{0}={G}_{\rm c} \ltimes (\hat{T}_{34}, {T}_{12})$.
We will now show that the four free parameters $\gamma$, $\kappa$, 
$\alpha_{1}$, $\alpha_{2}$,
can be adjusted such that the 
theory admits an ${\cal N}=(4,4)$ supersymmetric 
$AdS$ ground state, leaving only two free parameters that
correspond to the the radii of the two $S^{3}$ spheres.
Furthermore, expanding the action~(\ref{action})
around this ground state precisely reproduces
the spectrum of table~\ref{spin32}.

In order to show, that the Lagrangian~(\ref{action})
admits an $AdS$ ground state, we first have to check 
the condition (\ref{stationary}) equivalent to 
the existence of a stationary point of the scalar potential~(\ref{pot}).
For this in turn 
we have to compute the tensors $A_1$, $A_2$ and $A_3$
(\ref{A123}) from the $T$-tensor~(\ref{T})
evaluated at the ground state ${\cal V}=I$. 
At this point, the $T$-tensor coincides with the embedding tensor~(\ref{theta2}).
The only technical problem is the translation from 
$\Theta$~(\ref{theta2}) in the $SO(8,8)$ basis of appendix~\ref{A2}
into the $SO(16)$ basis of appendix~\ref{A1} in which the
tensors $A_1$, $A_2$ and $A_3$  are defined.

It follows from (\ref{theta2}) that
$\Theta$ is traceless, $\theta =0$, and moreover that 
all components of $\Theta$, which mix bosonic and spinorial parts, like 
$\Theta_{ab|\alpha\dot{\beta}}$, vanish.
As a consequence, the tensor 
$A_1$ is block-diagonal, and its explicit form turns out to be
 \bea\label{a1}
  A_1^{IJ}=\frac{1}{7}\left(\begin{array}{cc} 
  2\Theta_{\dot{\alpha}\gamma|\dot{\beta}\gamma}
  +\overline{\Gamma}^{\hat{a}\hat{b}}_{\dot{\alpha}\dot{\gamma}}
  \overline{\Gamma}^{\hat{c}\hat{d}}_{\dot{\beta}\dot{\gamma}}
  \Theta_{\hat{a}\hat{b}|\hat{c}\hat{d}}  & 0 \\ 0 & 
  2\Theta_{\dot{\gamma}\alpha|\dot{\gamma}\beta}
  +\Gamma^{ab}_{\alpha\gamma}\Gamma^{cd}_{\beta\gamma}\Theta_{ab|cd}
  \end{array}\right)\;,
 \eea
with $SO(8)$ $\Gamma$-matrices $\Gamma^{a}_{\alpha\dot{\beta}}$,
see appendix~\ref{A22} for details.
Similarly, $A_2$ and $A_3$ are also block-diagonal and can be written as 
 \bea\label{a2}
  A_2^{I\dot{A}}=-\frac{1}{7}\left(\begin{array}{cc}  
  2\Gamma^a_{\gamma\dot{\epsilon}}\Theta_{\dot{\alpha}\gamma|
  \beta\dot{\epsilon}}-\Gamma^{\hat{b}}_{\beta\dot{\gamma}}
  \overline{\Gamma}^{\hat{c}\hat{d}}_{\dot{\alpha}\dot{\gamma}}
  \Theta_{\hat{c}\hat{d}|a\hat{b}} & 0 \\
  0 & 2\Gamma^{\hat{a}}_{\delta\dot{\gamma}}\Theta_{\dot{\gamma}\alpha|\delta
  \dot{\beta}}+\Gamma^{cd}_{\alpha\gamma}\Gamma^b_{\gamma\dot{\beta}}
  \Theta_{cd|b\hat{a}} \end{array}\right),
 \eea
and
 \bea\label{a3}
   A_3^{\dot{A}\dot{B}}=\left(\begin{array}{cc} \delta^{\alpha\beta}
   \Theta_{a\hat{c}|b\hat{c}}+2\delta^{ab}\Theta_{\alpha\dot{\gamma}|
   \beta\dot{\gamma}}  & 0 \\ 
   0 & \delta^{\dot{\alpha}
   \dot{\beta}}\Theta_{c\hat{a}|c\hat{b}}+2\delta^{\hat{a}\hat{b}}
   \Theta_{\gamma\dot{\alpha}|\gamma\dot{\beta}}\end{array}\right).
 \eea
Using these tensors one can now check that the ground state condition 
(\ref{stationary}) is fulfilled if the parameters of (\ref{theta2}) satisfy
 \bea\label{groundstate}
  \kappa^{2}&=&16 \gamma^2\;.
 \eea
Moreover, using (\ref{pot}) the value of the scalar potential
at the ground state, i.e.~the cosmological constant, can be computed
and consistently comes out to be negative~$V=-g^{2}/2$,
i.e.~the $AdS$ length is given by $L_{0}=1/g$.

In the following, we will absorb $\kappa$ into the global coupling constant 
$g$ and set $\gamma=1/4$ in accordance with~(\ref{groundstate}).
As a result, there remains a two-parameter family of supergravities 
admitting an $AdS$ ground state. 
Let us now determine the number of unbroken supercharges in the ground state.
It is derived from the Killing spinor equations
 \bea\label{cond1}
  \delta\psi_{\mu}^I = D_{\mu}\epsilon^I+ig A_1^{IJ}\gamma_{\mu}
  \epsilon^J\equiv 0\;,
  \qquad
  \delta\chi^{\dot{A}}=gA_2^{I\dot{A}}\epsilon^I\equiv 0
  \;.
 \eea
As has been shown in~\cite{Nicolai:2000sc},
the number of solutions to~(\ref{cond1}) and thus the
number of preserved supersymmetries is given by the
number of eigenvalues $\alpha_{i}$ of the tensor $A_{1}^{IJ}$
with $|\alpha_{i}|gL_{0}=1/2$.
Computing these eigenvalues from the explicit form of~(\ref{a1}) we find
that the tensor $A_{1}^{IJ}$
may be diagonalized as
\bea
\label{eigenvalues}
A_{1}^{IJ} &=&
{\rm diag}\,
\Big\{
-\ft{3}{2}, -\ft{3}{2}, -\ft{3}{2}, -\ft{3}{2}, 
-\ft{1}{2},-\ft{1}{2},-\ft{1}{2},-\ft{1}{2},
\ft{1}{2},\ft{1}{2},\ft{1}{2},\ft{1}{2},
\ft{3}{2}, \ft{3}{2}, \ft{3}{2}, \ft{3}{2} 
\Big\} \;.
\eea
{}From this, we infer that the $AdS$ ground state of the theory
indeed preserves ${\cal N}=(4,4)$ supersymmetries, as expected.
The other eight gravitinos become
massive through a super-Higgs mechanism~\cite{deWit:2003ja, Hohm:2004rc}. 
This implies that due to the broken supersymmetries eight of the
spin-$1/2$ fermions 
 \bea
   \eta^I&\equiv& A_2^{I\dot{A}}\chi^{\dot{A}} \;,
   \label{gs}
 \eea
transform by a shift under supersymmetry and
act as Goldstone fermions that get eaten by the gravitino fields which in turn
become massive propagating spin-$3/2$ fields.  
With the relation
\bea
 \label{mass}
  |\Delta-1|= |m|\,L_{0} \;,
\eea
between the $AdS$ masses $m$ and conformal dimensions $\Delta$
of fermions and self-dual massive vectors
in three dimensions, (\ref{eigenvalues}) 
implies that the massive gravitinos correspond to
operators with conformal weights $(\ft12,2)$ and $(2,\ft12)$,
in precise agreement with the spectrum of table~\ref{spin32}.

To compute the physical masses for the spin-$1/2$ fermions,
we observe from (\ref{action}) that their mass matrix is given by
$gA_3^{\dot{A}\dot{B}}$, except for the eight eigenvalues that 
correspond to the Goldstone fermions~(\ref{gs}). 
From the explicit form (\ref{a3}) 
one computes the spin-1/2 
masses and verifies using~(\ref{mass})
that they coincide with those of table~\ref{spin32}.
Finally, we may check the mass spectrum for the spin-$1$ fields.
Their mass matrix is given 
by $\Theta_{AB}$, the projection of the embedding tensor
onto the non-compact part of the algebra~\cite{deWit:2003ja}. 
From (\ref{theta2}) one finds by explicit computation
for these eigenvalues $46$ non-vanishing values in precise accordance with 
table~\ref{spin32}.

Altogether, we have shown the existence of a new family of gauged maximally 
supersymmetric theories in $D=3$, which are parametrized by the two
free parameters $\alpha_1$ and $\alpha_2$ and the overall gauge coupling
constant $g$. 
These theories admit an ${\cal N}=(4,4)$ supersymmetric
$AdS_{3}$ ground state and linearizing the field equations around this 
ground state reproduces the correct spectrum of table~\ref{spin32}.
In particular, this spectrum does not depend on the particular
values of $\alpha_{1}$ and $\alpha_{2}$.
One may still wonder about the 
meaning of these two parameters. From the point of view 
of the Kaluza-Klein reduction the only relevant parameter 
is the ratio $\alpha$ of the two spheres radii, which enters the 
superalgebra~(\ref{supergroup}).
Let us thus compute the background isometry group by expanding 
the supersymmetry algebra
\bea
{}\{\,\delta_{\epsilon_{1}},\delta_{\epsilon_{2}} \,\}
&=& (\bar\epsilon_{1}^{I}\epsilon_{2}^{J}) 
\,{\cal V}^{\cM}{}_{{IJ}}\, \Theta_{{\cM\cN}}\,t^{\cN} + \dots
\;,
\label{bison}
\eea
around the ground state ${\cal V}=I$.
The conserved supercharges $\epsilon^{I}$ are the
eigenvectors of $A_{1}$ from (\ref{eigenvalues}) 
to the eigenvalues $\pm1/2$ where the different signs 
correspond to the split into
left and right supercharges according to~(\ref{supergroup}).
Correspondingly, the algebra~(\ref{bison}) 
splits into two parts, $L$ and $R$, with anticommutators
\bea
\label{susyclosure}
   \{ G_{-1/2\,L,R}^i,G_{1/2\,L,R}^j \}=
   4\left(\frac{1}{1+\alpha_{L,R}}\,\tau_{kl}^{+ij}\,J^{+kl}_{L,R}
   +\frac{\alpha_{L,R}}{1+\alpha_{L,R}}\,\tau_{kl}^{-ij}\,J^{-kl}_{L,R}\right)+\dots
   \;,
\eea
where $\tau_{kl}^{+ij}\equiv \delta_{kl}^{ij}\pm\ft12\epsilon_{ijkl}$
denote the projectors onto selfdual and anti-selfdual generators of 
$SO(4)_{L,R}$ corresponding to the split
$SO(4)=SO(3)^{+}\times SO(3)^{-}\,$.
This coincides with the anticommutators of the superalgebra
$D^{1}(2,1;\alpha_{{L,R}})$~\cite{Sevrin:1988ew}.
Specifically, we find the relation
\bea
   \alpha_{L}= \tan \alpha_1
   \;, \qquad 
   \alpha_{R} = \tan \alpha_2\;,
\eea
to the parameters (\ref{alpha12}) of the embedding tensor.
This shows that the three-parameter family of theories constructed
in this section exhibits the background isometry group
\bea
D^1(2,1;\alpha_{L})_{L} \times D^1(2,1;\alpha_{R})_{R}
\;.
\label{supergroup0}
\eea
The theories related to the Kaluza-Klein compactification
on $AdS_{3}\times S^{3}\times S^{3}$ are thus given 
by further restricting $\alpha_{L}=\alpha_{R}\equiv\alpha$
where this parameter corresponds to the ratio of radii of the two spheres.

Putting everything together, we have shown that the effective supergravity action
describing the field content of table~\ref{spin32} is given by the 
Lagrangian~(\ref{action}) with the following particular form
of the embedding tensor $\Theta_{\cM\cN}$
 \bea
 \label{thetatotal}
&&
    \Theta_{\f3_L^+,\f{\hat3}{}_L^+} =\Theta_{\f3_R^+,\f{\hat3}{}_R^+}=
    \frac{\alpha}{\sqrt{1\!+\!\alpha^{2}}}\;, \qquad
    \Theta_{\f3_L^-,\f{\hat3}{}_L^-}=
    \Theta_{\f3_R^-,\f{\hat3}{}_R^-}=\frac{1}{\sqrt{1\!+\!\alpha^{2}}}\;, \nonumber\\[1ex]
&&
   \Theta_{\f{1}_+,\f{1}_-}~=~
    \Theta_{\f{\hat3}{}_R^+,\f{\hat3}{}_R^+}~=~
    \Theta_{\f{\hat3}{}_R^-,\f{\hat3}{}_R^-}~=~
    -\Theta_{\f{\hat3}{}_L^+,\f{\hat3}{}_L^+}~=~
    -\Theta_{\f{\hat3}{}_L^-,\f{\hat3}{}_L^-}~=~\frac14 \;,  \nonumber\\[1ex]
&&
   \Theta_{{\f{16}}_{+}^{(1)},\f{16}^{(1)}_-}
    =-\frac{1}{16\sqrt{2}}\,\frac{1}{\sqrt{1\!+\!\alpha^{2}}}\;, \qquad
    \Theta_{{\f{16}}_{+}^{(1)}, 
    \f{16}^{(2)}_-}=-\frac{1}{16\sqrt{2}}\,\frac{\alpha}{\sqrt{1\!+\!\alpha^{2}}}\;. 
 \eea
We have verified that this tensor indeed represents a solution of
the algebraic consistency constraints~(\ref{conex}),~(\ref{quad}).
The resulting theory admits an ${\cal N}=(4,4)$ supersymmetric
$AdS_{3}$ ground state with background isometry group~(\ref{supergroup})
at which half of the 16 supersymmetries are spontaneously broken
and the spectrum of table~\ref{spin32} is reproduced via a supersymmetric
version of the Brout-Englert-Higgs effect.

\subsection{The scalar potential for the gauge group singlets}
\label{sec:pot}

We have identified the gauged supergravity theory,
whose broken phase describes the coupling of
the massive spin-$3/2$ multiplet $(\ft12,\ft12;\ft12,\ft12)_{\rm S}$
to the supergravity multiplet.
In particular, the scalar potential~(\ref{pot}) of the 
effective three-dimensional theory is completely
determined in terms of the embedding tensor~(\ref{thetatotal}).
In the holographic context this scalar potential
carries essential information about the boundary
conformal field theory, in particular 
about higher point correlation functions 
and about deformations and renormalization group flows.
Explicit computation of the full potential
is a highly nontrivial task, as it is a function
on the $128$-dimensional target space
$E_{8(8)}/SO(16)$. For concrete applications it is
often sufficient to evaluate this potential on
particular subsectors of the scalar manifold.

As an example, let us in this final section 
evaluate the potential on the gauge group singlets. From table~\ref{spin32}
we read off that there are two scalar fields that are singlets
under the $SO(4)_{L}\times SO(4)_{R}$ gauge group. 
Let us denote them by $\phi_{1}$ and $\phi_{2}$.
They are dual to a marginal and an irrelevant operator of 
conformal dimension $(1,1)$ and $(2,2)$, respectively.
In particular, 
the scalar $\phi_{1}$ corresponds to a modulus of the theory.
In order to determine the explicit dependence of the scalar
potential on these fields, we parametrize the scalar $E_{8(8)}$ 
matrix ${\cal V}$ as
\bea
  \mathcal{V}=\exp\big( \phi_1 X^{0\hat{0}}+\phi_2 X^{\bar0\hat{\bar0}} \big)
  \;,
\label{VV}
\eea
where $X^{0\hat{0}}$ and $X^{\bar0\hat{\bar0}}$ are the 
generators of the $SO(1,1)_{a}$ and $SO(1,1)_{b}$
of table~\ref{grading}, respectively.
The potential is obtained by computing with this parametrization 
the $T$-tensor from~(\ref{T}), (\ref{thetatotal}),
splitting it into the tensors $A_1$ and $A_2$ according to~(\ref{A123})
and inserting the result into~(\ref{pot}).

The computation is simplified by first transforming the two singlets
into a basis where their adjoint action is diagonal, such that
their exponentials can be easily computed and afterwards
transforming back to the $SO(16)$ basis of appendix~\ref{A1}.
It becomes now crucial that the embedding tensor $\Theta$ is 
invariant under $SO(1,1)_{a}$ and thus under the 
adjoint action of $X^{0\hat{0}}$ as we found in
section~\ref{constr} above. This implies, that the $T$-tensor~(\ref{T})
is in fact completely independent of $\phi_{1}$. In turn, neither the fermionic mass
terms nor the scalar potential carries an explicit dependence on $\phi_{1}$.
This scalar thus enters the theory only through its kinetic term,
the dual operator is truly marginal.

\begin{figure}
 \begin{center}
   \mbox{\includegraphics[width=0.7\textwidth]{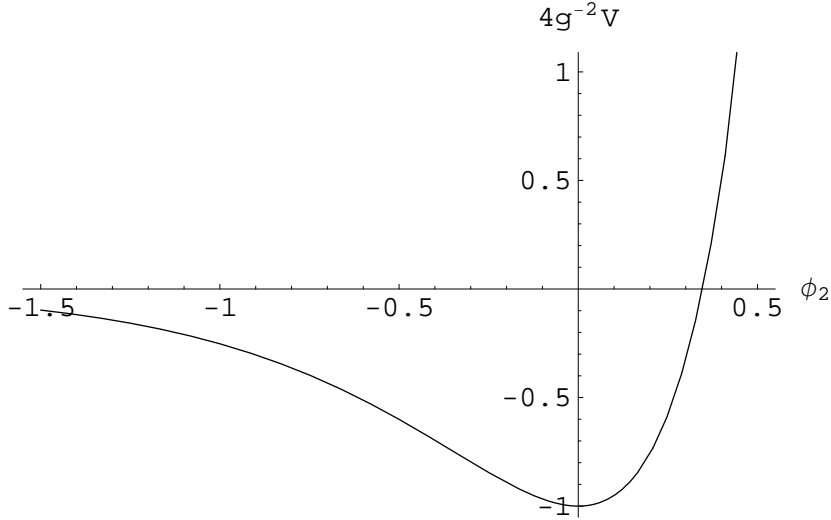}}
 \end{center}
\caption{The scalar potential for the gauge group singlets}\label{graph}
\end{figure}

The scalar potential (\ref{pot}) evaluated on the gauge group singlets
is finally given as a function of $\phi_{2}$ as
 \bea\label{potential}
  V(\phi_1,\phi_2)=\frac{g^2}{4}
  e^{2\phi_2}\big(-2 + e^{2\phi_2}\big).
 \eea
The profile is plotted in Figure~\ref{graph}.
Its explicit form shows that the theory has no other ground state 
which preserves the full $SO(4)\times SO(4)$ symmetry.

\section{Conclusions}
In this paper we have described the effective supergravity actions 
for the lowest massive Kaluza-Klein multiplets
in the Kaluza-Klein towers on $AdS_{3}\times S^{3}\times S^{3}$.
In particular, we have constructed the coupling of the massive
spin-$3/2$ multiplet $(\ft12,\ft12;\ft12,\ft12)_{\rm S}$
to the massless supergravity multiplet.
The corresponding three-dimensional
effective theory is the maximal gauged
supergravity~(\ref{action})
which is completely specified by the choice of the
embedding tensor~(\ref{thetatotal}).
We have verified that this tensor
indeed provides a solution to the algebraic
consistency constraints~(\ref{conex}), (\ref{quad}).
Specifically we have shown that these constraints
together with the requirement of the existence
of an $AdS_{3}$ ground state with the correct isometries
uniquely determine the theory
up to two constant parameters corresponding to the radii
of the two $S^{3}$ spheres.
The effective theory comes with a local gauge symmetry
$SO(4)_{L}\times SO(4)_{R}\ltimes (\hat{T}_{34}, {T}_{12})$
whose $46$-dimensional translational part is broken
at the $AdS$ ground state and gives rise to
the massive spin-$1$ fields of table~\ref{spin32}.
Similarly, half of the supersymmetries are spontaneously
broken at the ground state, and the full spectrum 
precisely reproduces 
the field content of table~\ref{spin32}.

The effective theory constructed in this paper 
offers various routes to shed further light
onto the holographic boundary CFT.
It will be of particular interest
to evaluate the scalar potential~(\ref{pot})
on different subsectors of the supergravity
scalars. This provides further insight into the 
structure of deformations of the boundary theory. 
For instance, the potential structure on
the scalars in the $(0,0;1,0)\oplus(0,0;0,1)$ representation
will encode further information
on the marginal deformations 
discussed in~\cite{Nishimura:2004ed}
that break the superconformal ${\cal N}=(4,4)$
symmetry down to~${\cal N}=(4,0)$.
Another interesting class of deformations of the boundary theory
are the relevant deformations by the dimension~$1$ operators
in the $(\ft12,\ft12;\ft12,\ft12)$ representation.

A natural question about the construction presented here
is the extension of the analysis
to the rest of the Kaluza-Klein spectrum.
Since the spin-$3/2$ multiplet is realized as a maximally
supersymmetric theory, it is impossible to 
couple further fields to it:
the scalar target space of the maximal theory
is the unique coset space
$E_{8(8)}/SO(16)$ whose $128$ degrees of freedom
match (and are exhausted by)
the bosonic content of the spin-$3/2$ multiplet. 
This is in contrast to the effective theories
on the $AdS_{3}\times S^{3}$ background in which 
the infinite Kaluza-Klein
tower of massive spin-$1$ multiplets
is realized as half maximal ${\cal N}=8$ supergravity 
theories~\cite{Nicolai:2003ux} whose scalar
target spaces are chosen from the series $SO(8,n)/(SO(8)\times SO(n))$
and may thus accommodate an arbitrarily large number
of degrees of freedom.
The Kaluza-Klein spectrum~(\ref{tower})
on the other hand contains already two copies
of the multiplet~$(\ft12,\ft12;\ft12,\ft12)_{\rm S}$
that together do not fit into the same maximal theory,
not to mention the infinite number of higher massive multiplets.
The effective theory describing the higher multiplets
in the Kaluza-Klein spectrum on $AdS_{3}\times S^{3}\times S^{3}$
will presumably require an extension of the construction
to a theory with an infinite number of supercharges
of which all but a finite number become massive
in a super Higgs effect at the ground state.
The usual no-go theorem that excludes the existence of theories 
with more than $32$ real supercharges
relies on the assumption that the theory admits an unbroken phase, 
such that the spectrum around this symmetric ground state
organizes into supermultiplets of the corresponding superalgebra.
With the theory constructed in this paper we have found
an example of a supersymmetric theory that 
has no symmetric phase but
only a half supersymmetric ground state.
Similarly, one could entertain the 
possibility of more than ${\cal N}=16$
supersymmetries in three dimensions 
which are necessarily broken at the ground state. 
Upon switching off the propagating matter,
the theories should reduce to the ${\cal N}$-extended
topological theory of~\cite{Achucarro:1987vz}.
We leave these questions for future study.

\subsection*{Acknowledgments}
We acknowledge helpful discussions with 
M.~Berg, S.~Gukov, H.~Jockers, and K.~Skenderis. 
This work is partly
supported by EU contracts
MRTN-CT-2004-503369 and MRTN-CT-2004-512194, 
and the DFG grant SA 1336/1-1. 
H.S. thanks the CERN theory division for hospitality 
when this work was being finalized.

\newpage

\begin{appendix}
\renewcommand{\theequation}{\Alph{section}.\arabic{equation}}

\section{Appendix: $E_{8(8)}$ from various angles}\label{A}

The maximal supergravity theories in three dimensions are organized
under the exceptional group $E_{8(8)}$. In particular, their scalar
sector is given by a coset space $\sigma$ model with target space
$E_{8(8)}/SO(16)$. In this appendix, we describe the Lie algebra
$\mathfrak{e}_{8(8)}$ in different decompositions relevant 
for the embedding of the gauge group and construction
of the embedding tensor in the main text.

\subsection{$E_{8(8)}$ in the $SO(16)$ basis}\label{A1}
\setcounter{equation}{0}

The $248$-dimensional Lie algebra of $E_{8(8)}$ may be characterized 
starting from its $120$-dimensional maximal compact subalgebra $\mathfrak{so}(16)$,
spanned by generators $X^{IJ}=X^{[IJ]}$ with commutators
 \bea
  [X^{IJ},X^{KL}]&=&\delta^{JK}X^{IL}-\delta^{IK}X^{JL}-\delta^{JL}X^{IK}
  +\delta^{IL}X^{JK}\;,
 \eea
where $I, J=1, \dots, 16$ denote $SO(16)$ vector indices. 
The $128$-dimensional non-compact part of $\mathfrak{e}_{8(8)}$ is spanned by 
generators $Y^A$ which transform in the fundamental spinorial representation of $SO(16)$,
i.e.~which satisfy commutators
 \bea
   [X^{IJ},Y^A]&=&-\ft12\,\Gamma_{AB}^{IJ}\,Y^B\;,
   \qquad
   [Y^A,Y^B]~=~\ft{1}{4}\,\Gamma_{AB}^{IJ}\,X^{IJ}\;.
 \eea
Here $A, B=1, \dots, 128$ label the spinor representation of $SO(16)$
and $\Gamma^{IJ}=\Gamma^{[I}\Gamma^{J]}$
denotes the antisymmetrized product of $SO(16)$ $\Gamma$-matrices.
Moreover, we use indices
$\dot{A}, \dot{B}=1, \dots, 128$ to label the conjugate spinor 
representation of $SO(16)$.
In the main text, indices $\mathcal{M}, \mathcal{N} =1, \dots, 248$
collectively label the full Lie algebra of $E_{8(8)}$, i.e.~
$\{t^{\cM}\} = \{X^{IJ},Y^{A}\}$ with
\bea
[\,t^{\cM},t^{\cN}\,] &=& f^{\cM\cN}_{\hspace{1.5em}\cK}\,t^{\cK}\;.
\eea
The Cartan-Killing form finally is given by 
 \bea\label{cartan}
  \eta^{\cM\cN}&=&\frac{1}{60}\,\text{tr}\,(t^{\cM}\,t^{\cN})~=~
  \frac{1}{60}\,
  f^{\cM\cK}_{\hspace{1.5em}\cL}\,f^{\cN\cL}_{\hspace{1.5em}\cK}\;.
 \eea

\subsection{$E_{8(8)}$ in the $SO(8,8)$ basis}\label{A2}

Alternatively, $\mathfrak{e}_{8(8)}$ may be built starting 
from its maximal
subalgebra $\mathfrak{so}(8,8)$ spanned by 120 generators $X^{{\tt IJ}}$
with commutators
 \bea
  [X^{{\tt IJ}},X^{\tt KL}]&=&\eta^{\tt JK}X^{\tt IL}
  -\eta^{\tt IK}X^{\tt JL}-\eta^{\tt JL}X^{\tt IK}
  +\eta^{\tt IL}X^{\tt JK}\;,
 \eea
where $\tt{I}, \tt{J}, \dots$ now denote vector indices of $SO(8,8)$
and $\eta_{\tt IJ}=\text{diag}(-1,\dots,-1,1,\dots,1)$ 
is the $SO(8,8)$ invariant tensor. 
Similarly to the above, the full $\mathfrak{e}_{8(8)}$
is obtained by adding 128 generators $\hat Q_{\tt A}$, ${\tt A} =1, \dots, 128$,
transforming in the spinor representation of $SO(8,8)$
\bea\label{rel1}
  [X^{\tt IJ},\hat Q_{\tt A}]&=&
  -\ft12 \Gamma^{{\tt IJ}}{}_{{\tt A}}{}^{{\tt B}}\,\hat Q_{\tt B}\;,
  \qquad
    [\hat Q_{{\tt A}},\hat Q_{{\tt B}}]~=~\ft14\,\eta_{{\tt IK}}\eta_{{\tt JL}}
  \eta_{{\tt BC}}\,
  \Gamma^{{\tt KL}}{}_{{\tt A}}{}^{{\tt C}}\,X^{{\tt IJ}}\;.
\eea
Here $\ga{{\tt IJ}}{{\tt B}}{{\tt A}}$ 
denote the (rescaled) $SO(8,8)$-generators in the spinor representation, i.e. 
 \bea
  \Gamma^{{\tt IJ}}{}_{{\tt A}}{}^{{\tt B}}
   =\ft12(\ga{{\tt I}}{\dot{{\tt C}}}{{\tt A}}\,\gaquer{{\tt J}}{{\tt B}}
   {\dot{{\tt C}}}-\ga{{\tt J}}{\dot{{\tt C}}}{{\tt A}}\,
  \gaquer{{\tt I}}{{\tt B}}{\dot{{\tt C}}})\;, 
 \eea
where the gamma matrices satisfy
 \bea 
 \label{gammaso88}
  \ga{{\tt I}}{\dot{{\tt C}}}{{\tt A}}\,\gaquer{{\tt J}}{{\tt B}}{\dot{{\tt C}}}+
  \ga{{\tt J}}{\dot{{\tt C}}}{{\tt A}}\,\gaquer{{\tt I}}{{\tt B}}{\dot{{\tt C}}} 
  =2\eta^{{\tt IJ}}\,\delta_{{\tt A}}^{\hspace{0.3em}{\tt B}}\;,
 \eea
with the transpose $\overline{\Gamma}$, and where $\tt{A}, \tt{B}, \dots$,
denote spinor indices and $\dot{\tt{A}}, \dot{\tt{B}}, \dots$, 
conjugate spinor indices of $SO(8,8)$.  
It is important to note that in contrast to the $SO(16)$ decomposition
described above, spinor indices in these equations are raised
and lowered not with the simple $\delta$-symbol but with the
corresponding $SO(8,8)$ invariant tensors $\eta_{\tt AB}$, $\eta_{\dot{\tt A}\dot{\tt B}}$
of indefinite signature (cf.~(\ref{eta88}) below).

\subsection{$E_{8(8)}$ in the $SO(8)\times SO(8)$ basis}\label{A22}

According to (\ref{embbb}), (\ref{16vs88}) 
in the main text, the two decompositions of sections~\ref{A1} and \ref{A2}
may be translated into each other upon further breaking down to  
$SO(8)_{L}\times SO(8)_{R}$.
To this end, we use the decomposition $SO(8,8)\rightarrow SO(8)_{L}\times SO(8)_{R}$
with
\bea\label{so88} 
   \f{16}_V&\rightarrow& (\f{8}_V,\f{1})\oplus (\f{1},\f{8}_V)\;, \nonumber\\  
   \f{128}_S&\rightarrow& (\f{8}_S,\f{8}_C)\oplus (\f{8}_C,\f{8}_S)\;,\qquad
   \f{128}_C~\rightarrow~ (\f{8}_S,\f{8}_S)\oplus (\f{8}_C,\f{8}_C)\;,
\eea
corresponding to the split of $SO(8,8)$ indices:
 \bea
  {\tt I} = (\li{a},b), \qquad {\tt A} = (\alpha\dot{\beta}\;,
  \dot{\gamma}\delta)\;, \qquad
  \dot{{\tt A}}=(\alpha \beta, \dot{\gamma}\dot{\delta})\;.
 \eea
Here, $\li{a}, \li{b}, \dots$ and $a, b, \dots$ denote vector indices for
the left and the right $SO(8)$ factor, respectively, while
$\alpha, \beta, \dots$ and $\dot{\alpha}, \dot{\beta}, \dots$ denote
spinor and conjugate spinor indices, respectively, for both $SO(8)$ factors. 
The invariant tensors
$\eta^{{\tt IJ}}$, $\eta_{{\tt AB}}$ and $\eta_{\dot{{\tt A}}\dot{{\tt B}}}$ 
in this $SO(8)$ notation take the form
 \begin{equation}\label{eta88}
  \begin{split} 
   \eta^{{\tt IJ}}=\left(\begin{array}{cc} -\delta^{\li{a}\li{b}} & 0 \\
   0 & \delta^{ab} \end{array}\right)\;, \qquad
   \eta_{{\tt AB}}&=\left(\begin{array}{cc} \eta_{\alpha\dot{\alpha},
   \beta\dot{\beta}} & 0 \\ 0 &
   \eta_{\dot{\alpha}\beta,\dot{\gamma}\delta}\end{array}\right)
   =\left(\begin{array}{cc} \delta_{\alpha\beta}\delta_{\dot{\alpha}
   \dot{\beta}} & 0 \\ 0 & -\delta_{\dot{\alpha}\dot{\gamma}}
   \delta_{\beta\delta}\end{array}\right)\;, \\[1ex]
   \eta_{\dot{{\tt A}}\dot{{\tt B}}}
   &=\left(\begin{array}{cc} \eta_{\alpha\beta,\gamma\delta} & 0 \\
   0 & \eta_{\dot{\alpha}\dot{\beta},\dot{\gamma}\dot{\delta}}
   \end{array}\right)
   =\left(\begin{array}{cc} \delta_{\alpha\gamma}\delta_{\beta\delta} & 0 \\
   0 & -\delta_{\dot{\alpha}\dot{\gamma}}
   \delta_{\dot{\beta}\dot{\delta}}\end{array}\right)\;.
  \end{split}
 \end{equation}
It is straightforward to verify, that the $SO(8,8)$ gamma matrices~(\ref{gammaso88}) can be
expressed in terms of the $SO(8)$ gamma matrices $\Gamma^a_{\alpha\dot{\gamma}}$ 
as (see also~\cite{Nicolai:1986jk})
 \bea
 \ga{a}{\delta\epsilon}{\beta\dot{\gamma}} &=& 
 \delta_{\beta\delta}
   \Gamma^a_{\epsilon\dot{\gamma}}
\;,
\qquad
 \ga{a}{\dot{\gamma}\dot{\delta}}{\dot{\alpha}\beta}
   ~=~
   -\delta_{\dot{\alpha}\dot{\gamma}}\Gamma_{\beta\dot{\delta}}^a\;, 
   \nonumber\\
   \ga{\li{a}}{\gamma\delta}{\dot{\alpha}\beta}
   &=&
   \delta_{\beta\delta}
   \Gamma_{\gamma\dot{\alpha}}^a
   \;, 
   \qquad 
   \ga{\li{a}}{\dot{\gamma}\dot{\delta}}{\alpha\dot{\beta}}~=~
   -\delta_{\dot{\beta}\dot{\delta}}
   \Gamma^a_{\alpha\dot{\gamma}}\;.
 \eea

With the results from the previous section,
$\mathfrak{e}_{8(8)}$ can now explicitly
be given in the $\mathfrak{so}(8)_{L} \oplus \mathfrak{so}(8)_{R}$ basis.
Generators split according to $\{X^{ab}, X^{\li{a}\li{b}},X^{a\li{b}},
\hat Q_{\alpha\dot{\beta}},\hat Q_{\dot{\gamma}\delta}\}$
with the commutation relations
\bea
 \label{e81}{}
   [X^{ab},X^{cd}]&=&\delta^{bc}X^{ad}
   -\delta^{ac}X^{bd}-\delta^{bd}X^{ac}
   +\delta^{ad}X^{bc}\;, \nonumber\\[.5ex] {}
   [X^{\li{a}\li{b}},X^{\li{c}\li{d}}]&=&
   -\delta^{\li{b}\li{c}}X^{\li{a}\li{d}}
   +\delta^{\li{a}\li{c}}X^{\li{b}\li{d}}
   +\delta^{\li{b}\li{d}}X^{\li{a}\li{c}}
   -\delta^{\li{a}\li{d}}X^{\li{b}\li{c}}\;, \nonumber\\[.5ex] {}
   [X^{ab},X^{c\li{d}}]&=&\delta^{bc}X^{a\li{d}}-\delta^{ac}X^{b\li{d}}\;, 
   \nonumber\\[.5ex] {}
   [X^{\li{a}\li{b}},X^{c\li{d}}]&=&\delta^{\li{a}\li{d}}X^{c\li{b}}
   -\delta^{\li{b}\li{d}}X^{c\li{a}}\;, \nonumber\\[.5ex] {}
   [X^{a\li{b}},X^{c\li{d}}]&=&\delta^{\li{b}
   \li{d}}X^{ac}-\delta^{ac}
   X^{\li{b}\li{d}}\;,\nonumber\\[1.5ex] {}
{}      [X^{ab},\hat Q_{\alpha\dot{\beta}}]&=&-\ft12\overline{\Gamma}{}^{[a}
   _{\dot{\beta}\epsilon}\Gamma_{\epsilon\dot{\delta}}^{b]}
   \hat Q_{\alpha\dot{\delta}}\;, 
   \qquad [X^{ab},\hat Q_{\dot{\alpha}\beta}]~=~-\ft12\Gamma_{\beta\dot{\zeta}}
   ^{[a}\overline{\Gamma}{}_{\dot{\zeta}\delta}^{b]}\hat Q_{\dot{\alpha}\delta}\;, 
   \nonumber\\[.5ex] {}
   [X^{\li{a}\li{b}},\hat Q_{\alpha\dot{\beta}}]&=&\ft12\Gamma_{\alpha 
   \dot{\epsilon}}^{[a}\overline{\Gamma}{}_{\dot{\epsilon}\gamma}^{b]} 
   \hat Q_{\gamma\dot{\beta}}\;, \qquad 
   [X^{\li{a}\li{b}},\hat Q_{\dot{\alpha}\beta}]~=~
   \ft12\overline{\Gamma}{}_{\dot{\alpha}\epsilon}^{[a}
   \Gamma_{\epsilon\dot{\gamma}}^{b]} \hat Q_{\dot{\gamma}\beta}\;, \nonumber\\[.5ex] {}
   [X^{a\li{b}},\hat Q_{\alpha\dot{\beta}}]&=&
   \ft12\Gamma_{\delta\dot{\beta}}^{a} \Gamma_{\alpha\dot{\gamma}}^{b}
   \hat Q_{\dot{\gamma}\delta}\;, 
   \qquad 
   [X^{a\li{b}}\;,\hat Q_{\dot{\alpha}\beta}]~=~
   \ft12\Gamma_{\beta\dot{\delta}}^{a} \Gamma_{\gamma\dot{\alpha}}^{b}
   \hat Q_{\gamma\dot{\delta}}\;,
   \nonumber\\[1.5ex] {}
      [\hat Q_{\alpha\dot{\alpha}},\hat Q_{\beta\dot{\beta}}]&=&\ft14\delta_{\alpha\beta}
   \overline{\Gamma}{}^{[a}_{\dot{\alpha}\gamma}\Gamma^{b]}_{\gamma\dot{\beta}}
   X^{ab}-\ft14\delta_{\dot{\alpha}\dot{\beta}}\Gamma^{[a}_
   {\alpha\dot{\gamma}}\overline{\Gamma}{}_{\dot{\gamma}\beta}^{b]}
   X^{\li{a}\li{b}}\;, \nonumber\\[.5ex] {}
   [\hat Q_{\alpha\dot{\alpha}},\hat Q_{\dot{\beta}\beta}]&=&
   -\ft12\Gamma_{\beta\dot{\alpha}}^{a}
   \Gamma_{\alpha\dot{\beta}}^{b} X^{a\li{b}}\;, \nonumber\\[.5ex] {}
   [\hat Q_{\dot{\alpha}\alpha},\hat Q_{\dot{\beta}\beta}]
   &=& \ft14\delta_{\alpha\beta}\overline{\Gamma}{}_{\dot{\alpha}\gamma}^{[a}
   \Gamma_{\gamma\dot{\beta}}
   ^{b]}X^{\li{a}\li{b}}
   -\ft14\delta_{\dot{\alpha}\dot{\beta}}\Gamma_{\alpha\dot{\gamma}}^{[a}
   \overline{\Gamma}{}_{\dot{\gamma}\beta}^{b]}X^{ab}\;.
 \eea

Moreover, the Cartan-Killing form (\ref{cartan}) can be computed 
in the $SO(8)\times SO(8)$ basis by 
use of this explicit form of the structure constants.
The result is
 \begin{equation}
\label{CK2}
  \begin{split}
   \eta^{ab,cd}&=-\delta^{[ab],[cd]}\;, \qquad \eta^{\hat{a}\hat{b},
   \hat{c}\hat{d}}=-\delta^{[\hat{a}\hat{b}],[\hat{c}\hat{d}]}\;, \qquad 
   \eta^{a\hat{b},c\hat{d}}=\delta^{ac}\delta^{\hat{b}\hat{d}}\;, \\
   \eta_{\alpha\dot{\beta},\gamma\dot{\delta}}&=\delta_{\alpha\gamma}
   \delta_{\dot{\beta}\dot{\delta}}\;, \qquad
   \eta_{\dot{\alpha}\beta,\dot{\gamma}\delta}=-\delta_
   {\dot{\alpha}\dot{\gamma}}\delta_{\beta\delta}\;,
  \end{split}
 \end{equation}
while all other components vanish.

Finally let us identify explicitly the $SO(16)$ subalgebra in this 
$SO(8,8)$ basis. With respect to the $SO(8)\times SO(8)$ decomposition
of $SO(16)$ in (\ref{so16deco}) the indices split according to
 \bea
  I=(\dot{\alpha},\beta)\;, \qquad A=(\alpha\dot{\beta},\hat{a}b)\;,
  \qquad \dot{A}=(\alpha a,\hat{b} \dot{\beta})\;.
 \eea
Correspondingly, the $SO(16)$ generators $X^{IJ}$ decompose into
$X^{\alpha\beta}$, $X^{\dot{\alpha}\dot{\beta}}$ 
and $X^{\alpha\dot{\beta}}$, and can be written in terms of the 
compact $E_{8(8)}$ generators by
 \bea\label{basechange}
  X^{\alpha\beta}=\ft{1}{2}\Gamma_{\alpha\beta}^{ab}X^{ab}\;, \qquad
  X^{\dot{\alpha}\dot{\beta}}=-\ft{1}{2}
  \Gamma_{\dot{\alpha}\dot{\beta}}^{\hat{a}\hat{b}}X^{\hat{a}\hat{b}}\;,
  \qquad
  X^{\alpha\dot{\beta}}= \hat Q_{\dot{\beta}\alpha}\;.
 \eea
That these satisfy the $SO(16)$ algebra can be verified explicitly 
by use of standard gamma matrix identities. The noncompact generators $Y^{A}$
are identified as 
\bea
Y^{\alpha\dot{\beta}} =  \hat{Q}_{\alpha\dot{\beta}} 
\;,\qquad
Y^{\hat{a}b} = X^{b\hat{a}} \;.
\eea
One immediately verifies that this split into compact and noncompact generators
is in agreement with the eigenvalues of the Cartan-Killing form~(\ref{CK2}).

\subsection{$E_{8(8)}$ in the $SO(4)\times SO(4)$ basis}\label{A3}

To explicitly describe the embedding of the gauge group 
$G_{0}={G}_{\rm c} \ltimes (\hat{T}_{34}, {T}_{12})$
described in section~\ref{embed}, we finally need the decomposition
of $E_{8(8)}$ under the $SO(4)_{L}\times SO(4)_{R}$ from~(\ref{so34}).
This is obtained from the previous section
upon further decomposition according 
to~(\ref{embedding33}), (\ref{so4embed}).
In $SO(8)_{R}$ indices $a$, $\alpha$, $\dot{\alpha}$, this corresponds to the splits
 \bea\label{indexsplit}
  a=([ij],0,\bar0)\;, \qquad \alpha =(i,j)\;, \qquad 
  \dot{\alpha}=(i,j)\;,
 \eea
and similarly for $SO(8)_{L}$. Here, $i, j, \dots$ denote $SO(4)$ vector indices.
The $SO(8)$ gamma matrices can then be expressed
in terms of the invariant $SO(4)$ tensors $\delta^{ij}$ and $\varepsilon^{ijkl}$ as
 \bea\label{gamma}
  \Gamma^{ij}=\left(\begin{array}{cc} \varepsilon^{ij} & 2\delta^{ij} \\
  -2\delta^{ij} & \varepsilon^{ij}\end{array}\right)\;, \qquad
  \Gamma^{0}=\left(\begin{array}{cc} \textbf{1} & 0 \\
  0 & -\textbf{1}\end{array}\right)\;, \qquad
  \Gamma^{\bar0}=\left(\begin{array}{cc} 0 & \textbf{1} \\
  \textbf{1} & 0 \end{array}\right)\;,
 \eea
 with the $4 \times 4$ matrices
  \bea
 {\bf 1}_{kl} &=& \delta_{kl}\;,\qquad
  (\varepsilon^{ij})_{kl}~=~\varepsilon^{ijkl}\;, \qquad 
  (\delta^{ij})_{kl}~=~\delta^{ij}_{kl}~=~
  \delta^{i[k}\delta^{l]j}\;.
 \eea
It is straightforward to check that the matrices~(\ref{gamma}) satisfy the standard
Clifford algebra, making use of the relations
 \begin{equation}
  \begin{split}
   \delta^{ij}(\delta^{mn})^t 
+   \delta^{mn}(\delta^{ij})^t 
+ \varepsilon^{ij}(\varepsilon^{mn})^t
+ \varepsilon^{mn}(\varepsilon^{ij})^t
   &~=~2\delta^{ij,mn}\,\textbf{1}, \\
   \varepsilon^{ij}(\delta^{mn})^t+\varepsilon^{mn}(\delta^{ij})^t
    - \delta^{ij}(\varepsilon^{mn})^t- \delta^{mn}(\varepsilon^{ij})^t&~=~0,
  \end{split}
 \end{equation}   
which can be proved using the identity $\varepsilon^{[ijkl}\delta^{m]}_n=0\,$.
Next we have to decompose these $\Gamma$-matrices into selfdual and
anti-selfdual parts, corresponding to (\ref{indexsplit}),
 \bea
   \Gamma^{ij}_{\pm}=\frac{1}{\sqrt{2}}
   \big(\Gamma^{ij}\pm\frac{1}{2}\varepsilon^{ijkl}\Gamma^{kl}\big),
 \eea
such that $\tilde{\Gamma}^{ij}_{\pm}
:=\frac{1}{2}\varepsilon^{ijkl}\Gamma^{kl}_{\pm}=\pm\Gamma^{ij}_{\pm}$.
Inserting the representation (\ref{gamma}) of $\Gamma$-matrices
into the structure constants in~(\ref{e81}) eventually yields the 
decomposition of 
$\mathfrak{e}_{8(8)}$ 
in the $\mathfrak{so}(4)_{L} \oplus \mathfrak{so}(4)_{R}$ basis.

\end{appendix}
    
%\bibliographystyle{Jopt2}
%\bibliography{refs}

{\small
\providecommand{\href}[2]{#2}\begingroup\raggedright\endgroup
}

\end{document}